\documentstyle[aps,prb]{revtex}

\input{epsf}

\def\E{{\cal E}}

\def\op#1{{\hat{#1}}}
\def\bm#1{\mbox{\boldmath$#1$\unboldmath}}
\def\rem#1{}
\def\ddd{d}
\def\bbalpha{{\bm\alpha}}

\begin{document}
\draft 

\title{Superconductivity in Ultrasmall Metallic Grains} 

\author{Fabian Braun and Jan von Delft} 

\address{Institut f\"ur Theoretische Festk\"orperphysik, Universit\"at
  Karlsruhe, 76128 Karlsruhe, Germany}

\twocolumn[\hsize\textwidth\columnwidth\hsize\csname%
@twocolumnfalse\endcsname%
\date{January 16, 1998}%
\maketitle%
  
\begin{abstract} 
  We develop a theory of superconductivity in ultrasmall (nm-scale)
  metallic grains having a discrete electronic eigenspectrum with mean
  level spacing $\ddd\simeq \tilde\Delta$ (bulk gap). The theory is
  based on calculating the eigenspectrum using a generalized BCS
  variational approach, whose (qualitative) applicability has been
  extensively demonstrated in studies of pairing correlations in
  nuclear physics. We discuss how conventional mean field theory
  breaks down with decreasing sample size, how the so-called {\em
    blocking effect} (the blocking of pair-scattering by unpaired
  electrons) weakens pairing correlations in states with non-zero
  total spin (thus generalizing a parity effect discussed previously),
  and how this affects the discrete eigenspectrum's behavior in a
  magnetic field, which favors non-zero total spin.  In ultrasmall
  grains, spin magnetism dominates orbital magnetism, just as in thin
  films in a parallel field; but whereas in the latter the
  magnetic-field induced transition to a normal state is known to be
  first-order, we show that in ultrasmall grains it is softened by
  finite size effects.  Our calculations qualitatively reproduce the
  magnetic-field dependent tunneling spectra for individual aluminum
  grains measured recently by Ralph, Black and Tinkham \mbox{[}Phys.\ 
  Rev.\ Lett.\ {\bf 78}, 4087 (1997)\mbox{]}.  We argue that
  previously-discussed parity effects for the odd-even ground state
  energy difference are presently not observable for experimental
  reasons, and propose an analogous parity effect for the
  pair-breaking energy that should be observable provided that the
  grain size can be controlled sufficiently well.  Finally,
  experimental evidence is pointed out that the dominant role played
  by time-reversed pairs of states, well-established in bulk and in
  dirty superconductors, persists also in ultrasmall grains.
\end{abstract}

\vskip0.0pc]    

\section{Introduction}

What happens to superconductivity when the sample is made very, very small?
Anderson \cite{Anderson-59} addressed this question already in 1959: he argued
that if the sample is so small that its electronic eigenspectrum becomes
discrete, with a mean level spacing $\ddd = 1 / {\cal N} (\varepsilon_F) \sim
1 / \mbox{Vol}$, ``superconductivity would no longer be possible'' when $\ddd$
becomes larger than the bulk gap $\tilde \Delta$. Heuristically, this is
obvious (see Fig.~\ref{fig:v2u2} below): $\tilde \Delta / \ddd$ is the number
of free-electron states that pair-correlate (those with energies within
$\tilde \Delta $ of $\varepsilon_F$), i.e. the ``number of Cooper pairs'' in
the system; when this becomes ${\stackrel{\scriptscriptstyle
    <}{\scriptscriptstyle \sim}} 1$, it clearly no longer makes sense to call
the system ``superconducting''.

Giaver and Zeller\cite{GiaverZeller-68,ZellerGiaver-69} were among the first
to probe Anderson's criterion experimentally: studying tunneling through
granular thin films containing electrically insulated Sn grains, they
demonstrated the existence of an energy gap for grain sizes right down to the
critical size estimated by Anderson (radii of $25$\AA\ in this case), but were
unable to prove that smaller particles are always normal.  Their concluding
comments are remarkably perspicuous:\cite{ZellerGiaver-69} ``There can be no
doubt, however, that in this size region the bulk theory of superconductivity
loses its meaning. As a matter of fact, perhaps we should not even regard the
particles as metallic because the energy-level spacing is large compared to $k
T$ and because there are very few electrons at the Fermi surface.  The
question of the lower size limit for superconductivity is, therefore, strongly
correlated with the definition of superconductivity itself.''

These remarks indicate succinctly why the study of superconductivity near its
lower size limit is of fundamental interest: the conventional bulk BCS
approach is not directly applicable, 
and some basic elements of the theory need to be
rethought, with the role of level discreteness demanding special attention.

First steps in this direction were taken by Strongin {\em et
  al.}\cite{Strongin-70} and by M\"uhlschlegel {\em et
  al.}\cite{Muehlschlegel-72}, who calculated the thermodynamic properties of
small superconducting grains.  However, since experiments at the time were
limited to studying ensembles of small grains (e.g.\ granular films), there
was no experimental incentive to develop a more detailed theory for an {\em
  individual}\/ ultrasmall superconducting grain, whose eigenspectrum, for
example, would be expected to reveal very directly the interplay between level
discreteness and pairing correlations.

This changed dramatically in 1995, when Ralph, Black and Tinkham (RBT)
\cite{Ralph-95} succeeded in constructing a single-electron transistor (SET)
whose island was an ultrasmall metallic grain: by studying the tunneling
current through the device, they achieved the first measurement of the
discrete eigenspectrum of a single grain. This enabled them to probe the
effects of spin-orbit scattering,\cite{Black-96,Ralph-96B} non-equilibrium
excitations\cite{Ralph-97} and superconductivity,\cite{Black-96,Ralph-97}
which manifests itself through the presence (absence) of a substantial
spectral gap in grains with an even (odd) number of electrons.

RBT's work stimulated several theoretical investigations.  Besides discussing
non-equilibrium effects,\cite{Agam-97,AgamAleiner-97} these focused mainly on
superconductivity,\cite{vonDelft-96,Smith-96,Matveev-97,Braun-97,Bahcall-priv}
and revealed that the breakdown of pairing correlations with decreasing grain
size predicted by Anderson harbors some surprises when scrutinized in more
detail: von Delft {\em et al.}\cite{vonDelft-96} showed that this breakdown is
affected by the {\em parity}\/ ($p$) of the number of electrons on the grain:
using parity-projected mean-field theory\cite{Janko-94,Golubev-94} 
 and variational methods and assuming
uniformly spaced electron levels, they solved the parity-dependent gap
equation for the even or odd ground state pairing parameters $\Delta_e$ or
$\Delta_o$ as function of $\ddd$ (using methods adapted from Strongin {\em et
  al.}\cite{Strongin-70}), and found that $\Delta_o (\ddd) < \Delta_e (\ddd)$,
i.e.\ ground state pairing correlations break down sooner with increasing
$\ddd$ in an odd than an even grain (the difference becoming significant for
$\ddd \simeq \tilde \Delta$).  This is due to the so-called {\em blocking
  effect}:\cite{Soloviev-61} the odd grain always has one unpaired electron,
which blocks pair-scattering of other pairs and thereby weakens pairing
correlations.  Smith and Ambegaokar\cite{Smith-96} showed that this parity
effect holds also for a random distribution of level spacings (as also
anticipated by Blanter\cite{Blanter-96}), and Matveev and
Larkin\cite{Matveev-97} investigated a parity effect occuring in the limit
$\ddd \gg \tilde \Delta$.

The $\Delta_o< \Delta_e$ parity effect has an obvious generalization, studied
by Braun {\em et al.}\/\cite{Braun-97} using a generalized BCS variational
approach due to Soloviev:\cite{Soloviev-61} {\em any} state with non-zero spin
$s$ (not just the odd ground state) experiences a significant reduction in
pairing correlations, since at least $2s$ electrons are unpaired, leading to
an enhanced blocking effect ($\Delta_s < \Delta_{s'}$ if $s > s'$).  The
latter's consequences can be observed in the magnetic-field dependence of SET
tunneling spectra, since a magnetic field favors states with non-zero spin and
consequent enhanced blocking effect.  In ultrasmall grains, spin magnetism
dominates orbital magnetism, just as in thin films in a parallel
field;\cite{Meservey-94} but whereas in the latter the magnetic-field induced
transition to a normal state is known to be first-order, Braun {\em et al.}\/
showed that in ultrasmall grains the transition is softened due to finite size
effects.  Moreover, they argued that some of RBT's grains fall in a region of
``minimal superconductivity'', in which pairing correlations measurably exist
at $H=0$, but are so weak that they may be destroyed by the breaking of a
single pair (since the number of electron pairs that take part in the
formation of a correlated state becomes of order one for $\ddd \simeq \tilde
\Delta $).

In the present paper we elaborate the methods used and results found by Braun
{\em et al.}  in Ref.~\onlinecite{Braun-97} and present a detailed theory of
superconductivity in ultrasmall grains. Our discussion can be divided into two
parts: in the first (sections~\ref{meaningofsc} and \ref{generalBCS}), we
consider an isolated ultrasmall grain and (a) define when and in what sense it
can be called ``superconducting''; (b) use a generalized BCS variational
approach to calculate the eigenenergies of various variational eigenstates of
general spin $| s \rangle$, which illustrates the break-down of mean-field
theory; and (c) discuss how an increasing magnetic field induces a transition
to a normal paramagnetic state.  In the second part
(section~\ref{observables}), we consider the grain coupled to leads as in
RBT's SET experiments and discuss observable quantities: (a) We calculate
theoretical tunneling spectra of the RBT type, finding qualitative agreement
with RBT's measurements; (b) point out that the above-mentioned ground state
energy parity effect can presently not be observed, and propose an analogous
pair-breaking energy parity effect that should be observable in experiments of
the present kind; and (c) explain how RBT's experiments give direct evidence
for the dominance of time-reversed pairing, at least for small fields
(implying that the sufficiency of using only a reduced BCS-Hamiltonian,
well-established for bulk systems and dirty superconductors, holds for
ultrasmall grains, too).

\section{Pairing Correlations at fixed Particle Number}
\label{meaningofsc}

The discrete energies measured in RBT's experiments essentially correspond to
the eigenspectrum of a grain with {\em fixed electron number $N$} (for reasons
explained in detail in section \ref{expdetails}).  In this and the next
section, we therefore consider an ultrasmall grain {\em completely isolated}\/
from the rest of the world, e.g.\ by infinitely thick oxide barriers.

When considering a truly isolated superconductor (another example
would be a superconductor levitating in a magnetic field due to the
Meissner effect) one needs to address the question: How is one to
incorporate the fixed-$N$ condition into BCS theory, and how important
is it to do so?  Although this issue is well understood and was
discussed at length in the early days of BCS theory, in particular in
its application to pairing correlations in nuclei \cite[p.
439]{RingSchuck-80}, for pedagogical reasons the arguments are worth
recapitulating in the present context.  We shall first recall that the
notion of pair-mixing \cite{vonDelft-96} that lies at the heart of BCS
theory is by no means inherently grand-canonical and can easily be
formulated in canonical language, then summarize what has been learned
in nuclear physics about fixed-$N$ projection techniques, and finally
conclude that for present purposes, standard grand-canonical BCS
theory should be sufficient.  Readers familiar with the relevant
arguments may want to skip this section.

\subsection{Canonical Description of Pair-Mixing}

Conventional BCS theory gives a grand-canonical description of the pairing
correlations induced by the presence of an attractive pairing interaction such
as the reduced BCS interaction
\begin{equation} 
  \label{eq:hamilton-1}
  H_{red} = - \sum_{j j'} V c^\dagger_{j+} c^\dagger_{j-}  c_{j'-} c_{j' +} \; 
  \quad (\mbox{with}\; V > 0).
\end{equation}
(The $c_{j\pm}$ are electron destruction operators for the single-particle 
states $|j,\pm\rangle$, taken to be time-reversed copies of each other,
with energies $\varepsilon_{j \pm}$.)
The theory employs a grand-canonical ensemble,
formulated on a Fock space of states in which the total particle number $N$ is
not fixed, as illustrated by BCS's variational ground state Ansatz
\begin{eqnarray}
  \label{eq:BCSground}
    |BCS \rangle = \prod_j 
    (u_j + v_j c^\dagger_{j+}c^\dagger_{j-})\,|\mbox{Vac}\rangle \; 
    \quad (u^2_j + v^2_j = 1) .
\end{eqnarray}            
This is not an eigenstate of the number operator $\hat N = \sum_{j \sigma}
c^\dagger_{j\sigma } c_{j\sigma} $ and its particle number is fixed only on
the average by the condition $ \langle BCS|\hat N |BCS \rangle = N$, which
determines the grand-canonical chemical potential $\mu$.  Likewise, the
commonly used definition
\begin{equation}
  \label{eq:BCS-gap}
  \Delta_{BCS} = V \sum_j \langle c_{j+}c_{j-} \rangle \; 
\end{equation}
for the superconducting order parameter only makes sense in a grand-canonical
ensemble, since it would trivially give zero when evaluated in a canonical
ensemble, formulated on a strictly fixed-$N$ Hilbert space of states.

A theory of strictly fixed-$N$ superconductivity must therefore 
entail modifications of conventional BCS theory.  In particular, a construction
different from $\Delta_{BCS}$ is needed for the order parameter, which we
shall henceforth call ``pairing parameter'', since ``order parameter'' carries
the connotation of a phase transition, which would require the thermodynamic
limit $N\to \infty$.  The pairing parameter should capture in a canonical
framework BCS's essential insight about the nature of the superconducting
ground state: an attractive pairing interaction such as $H_{red}$ will induce
pairing correlations in the ground state that involve {\em pair-mixing}\/
across $\varepsilon_F$ (see also Ref.~\onlinecite{vonDelft-96}), i.e.\ a
non-zero amplitude to find a pair of time-reversed states occupied above
$\varepsilon_F$ or empty below $\varepsilon_F$. BCS chose to express this
insight through the Ansatz (\ref{eq:BCSground}), which allows $v_j \neq 0$ for
$\varepsilon_j > \varepsilon_F$ and $u_j \neq 0$ for $\varepsilon_j <
\varepsilon_F$. It should be appreciated, however (and is made clear on
p.~1180 of their original paper \cite{BCS-57}), that they chose a {\em
  grand-canonical}\/ construction purely for calculational convenience (the
trick of using commuting products in (\ref{eq:BCSground}) makes it brilliantly
easy to determine the variational parameters $u_j$, $v_j$), and proposed
themselves to use its projection to fixed $N$, $|BCS\rangle_N$, as the actual
ground state.

Since $[H_{red}, \hat N ] = 0$, one would expect that the essence of BCS
theory, namely the presence of pair-mixing and the reason why it occurs, can
also be formulated in a canonically meaningful way. Indeed, this is easy:
pair-mixing is present if the amplitude $\bar v_j \equiv \langle
c^\dagger_{j+} c^\dagger_{j-} c_{j-} c_{j+} \rangle^{1/2}$ to find a pair of
states occupied is non-zero also for $\varepsilon_j > \varepsilon_F$, and the
amplitude $\bar u_j \equiv \langle c_{j-} c_{j+} c^\dagger_{j+} c^\dagger_{j-}
\rangle^{1/2}$ to find a pair of states empty is non-zero also for
$\varepsilon_j < \varepsilon_F$ (the bars indicate that the $\bar u_j$ and
$\bar v_j$ defined here differ in general from the $u_j$ and $v_j$ used by
BCS; note, though, that the former reduce to the latter if evaluated using
$|BCS \rangle$).  The intuitive reason why $H_{red}$ induces pair-mixing in
the exact ground states $|G \rangle$ despite the kinetic energy cost incurred
by shifting pairing amplitude from below to above $\varepsilon_F$, is that
this frees up phase space for pair-scattering, thus lowering the ground state
expectation value of $H_{red}$: in $\langle G | H_{red} | G \rangle$, the $j
j'$ term can be non-zero only if {\em both}\/ $c^\dagger_{j+} c^\dagger_{j-}
c_{j'-} c_{j' +} | G \rangle \neq 0$, implying $(\bar v_{j'})_G \neq 0$ and
$(\bar u_j)_G \neq 0$, {\em and also}\/ $\langle G | c^\dagger_{j+}
c^\dagger_{j-} c_{j'-} c_{j' +} \neq 0$, implying $(\bar v_j)_G \neq 0$ and
$(\bar u_{j'})_G \neq 0$.  By pair-mixing, the system can arrange for a
significant number of states to simultaneously have both $(\bar v_j)_G \neq 0$
and $(\bar u_j)_G \neq 0$; this turns out to lower the ground state energy
sufficiently through $\langle G | H_{red} | G \rangle$ that the kinetic energy
cost of pair-mixing is more than compensated. Furthermore, an excitation that
disrupts pairing correlations in the ground state by ``breaking up a pair''
will cost a finite amount of energy by {\em blocking}\/ pair-scattering
involving that pair.  For example, the energy cost of having $|j +\rangle$
definitely occupied ($\bar u_j = 0$) and $|j -\rangle$ definitely empty ($\bar
v_j = 0$) is \label{p:blocking}
\[
\varepsilon_j (1- \langle G | \sum_\sigma c^\dagger_{j \sigma} c_{j \sigma}|G
\rangle) + V \langle G | c^\dagger_{j +} c^\dagger_{j-} \!\! \sum_{j' \neq j}
c_{j'-} c_{j'+} |G\rangle,
\]
in which the restricted sum reflects the blocking of scattering involving
the $j$-th pair.  When evaluated using $|BCS\rangle$, this quantity reduces to
$\varepsilon_j (1 - 2 v_j^2) + u_j v_j \Delta_{BCS} = [ \varepsilon_j^2 +
\Delta_{BCS}^2]^{1/2}$, which is the well-known quasi-particle energy of the
state $\gamma^\dagger_{j+} |BCS\rangle$.

The above simple arguments illustrate that there is nothing inherently
grand-canonical about pair-mixing.  Indeed, at least two natural ways suggest
themselves to measure its strength in a canonically meaningful way, using for
instance the pairing parameter $\bar \Delta \equiv V \sum_j \bar u_j \bar v_j$
proposed in Ref.~\cite{vonDelft-96}, or one proposed by Ralph
\cite{Ralph-priv}:
\begin{equation}
\label{pairingparameter}
       \label{eq:v2u2}
    \bar \Delta' \equiv V \sum_j \left[
\langle c^\dagger_{j+}c_{j+} c^\dagger_{j-}c_{j-}\rangle -
  \langle c^\dagger_{j+}c_{j+} \rangle \langle c^\dagger_{j-}c_{j-}
  \rangle \right]^{1/2} \!\! .
\end{equation}
Both $\bar \Delta$ and $\bar \Delta'$ were constructed such that they reduce,
as is desirable, to the same result as $\Delta_{BCS}$ when each is evaluated
using $|BCS \rangle$ with real coefficients $u_j, v_j$, namely $V \sum_j u_j
v_j$.  An appealing feature of $\bar \Delta'$ is that by subtracting out $
\langle c^\dagger_{j+}c_{j+} \rangle \langle c^\dagger_{j-}c_{j-} \rangle$, it
transparently emphasizes the {\em pairing}\/ nature of superconducting
correlations, i.e.\ the fact that if $|j+ \rangle$ is empty (or filled), so is
$|j- \rangle$: $\bar \Delta'$ will be very small if the occupation of $|j +
\rangle$ is uncorrelated with that of $|j - \rangle$, as it is in a normal
Fermi liquid.  The overall behavior (as function of energy $\varepsilon_j$) of
the summands in both $\bar \Delta$ and $\bar \Delta'$ will be similar to that
of $u_j v_j$ (though not identical to $u_j v_j$ or to each other; a
quantitative evaluation of the differences, which increase with increasing
$\ddd / \tilde \Delta$, requires an honest canonical
calculation\cite{Braun-98}). $u_j v_j$ is shown in Fig.~\ref{fig:v2u2}(a),
which illustrates that pair-mixing correlations are strongest within a region
of width $\Delta_{BCS}$.  In this paper, we shall call a system
``pair-correlated'' if $\bar \Delta'$ is a significant fraction of its bulk
value $\tilde \Delta$ (say, somewhat arbitrarily, at least 25\%), and regard
this as being synonymous with ``superconducting''.

\subsection{On the breaking of Gauge Symmetry}

In some discussions of conventional BCS theory the defining feature of
superconductivity is taken to be the breaking of gauge symmetry by the order
parameter.  This concept is illustrated by the BCS order parameter $
\Delta_{BCS}$ of Eq.~(\ref{eq:BCS-gap}): if non-zero, it has a definite phase
and is not gauge-invariant (under $c_{j\sigma} \to e^{i \phi} c_{j\sigma}$, it
changes to $ e^{i 2 \phi} \Delta_{BCS}$).  Note, though, that this point of
view cannot be carried over to fixed-$N$ systems.  Firstly, these trivially
have $\Delta_{BCS}=0$, and secondly and more fundamentally, the breaking of
gauge symmetry necessarily presupposes a grand-canonical ensemble: since phase
and particle number are quantum-mechanically conjugate variables, formal
considerations dictate that the order parameter acquire a definite phase only
if the particle number is allowed to fluctuate, i.e.\ in a grand-canonical
ensemble.

Of course, in certain experimental situations where $N$ manifestly {\em
  does}\/ fluctuate, such as the celebrated Josephson effect of two
superconductors connected by a tunnel junction, their order parameters {\em
  do}\/ acquire definite phases, and their phase difference is a measurable
quantity. However, for a truly isolated superconductor with fixed $N$ the
``phase of the order parameter'' is {\em not}\/ observable, and the concept of
gauge symmetry breaking through an order parameter with a definite phase
ceases to be useful.  Indeed, the canonically meaningful pairing parameters
$\bar \Delta$ and $\bar \Delta'$ defined above are manifestly gauge-invariant.

\subsection{Fixed-$N$ Projections} 
\label{sec:fixedN}

It is easy to construct a variational ground state exhibiting pair-mixing {\em
  and}\/ having definite particle number, by simply projecting $|BCS \rangle$
to fixed $N$, as suggested by BCS \cite{BCS-57}. This can be achieved by the
projection integral
\begin{eqnarray}
  \label{eq:BCSground-N}
    |BCS \rangle_N \equiv \! \int_0^{2 \pi} \!\! \!\! \! \! d \phi\, 
    e^{-i \phi N} \! \prod_j 
    (u_j \! + \! e^{2 i \phi}
    v_j c^\dagger_{j+}c^\dagger_{j-})\,|\mbox{Vac}\rangle,
\end{eqnarray}      
whose randomization of the phases of the $v_j$'s illustrates, incidentally,
why gauge invariance is not broken at fixed $N$.

This and related fixed-$N$ projections were studied in great detail in nuclear
physics, with the aim of variationally calculating nuclear excitation spectra
for finite nuclei ($N \le 240$) exhibiting pairing correlations (Ring and
Schuck provide an excellent review of the extensive literature, see chapter 11
of Ref.~\onlinecite{RingSchuck-80}; a recent reference is
\onlinecite{Balain-97}). The simplest approach is called ``projection after
variation'': the unprojected expectation value $\langle BCS | H | BCS \rangle$
is minimized with respect to the variational parameters $\{ v_j \}$, which
thus have their standard BCS values $v_j^2 = {1 \over 2} [ 1 - \varepsilon_j /
(\varepsilon_j^2 + \Delta^2_{BCS})^{1/2}]$, but then these are inserted into
$|BCS \rangle_N$ and expectation values evaluated with the latter instead of
$|BCS \rangle$.  This elimination of ``wrong-$N$'' states after variation
turns out to lower the ground state energy relative to the unprojected case
(by a few percent in nuclei) and thus improves the trial wave-function.
Further improvements are possible using the more sophisticated ``projection
before variation'' strategy, where the projected expectation value
${}_N\langle BCS | H | BCS \rangle_N$ is minimized with respect to the $\{ v_j
\}$. However, these then no longer have the simple BCS form, but instead are
determined through a set of {\em coupled}\/ relations, each involving all the
other $v_{j}'$s, that have to be solved numerically.  The corrections $\delta
v_j$ to the BCS pair-occupation amplitudes so produced further lower the
ground state energy relative to projection after variation (but only by tenths
of a percent).

Extensive applications of such and related approaches in nuclear physics have
led to the following conclusions: For reasonably small $N$, as in nuclei, the
explicit implementation of projection techniques is tractable, though
cumbersome.  For very large $N$ they become intractable, but also unnecessary,
since their corrections can be shown to vanish as $N^{-1/2}$.  However, even
in nuclei the corrections to unprojected BCS theory are small (a few percent)
in most cases, the only exception being very large couplings $V \ge \ddd$.
Thus, in most cases fixed-$N$ systems can perfectly adequately be described by
BCS's grand-canonical wave function.  Its $N$-indefiniteness (and the
associated breaking of gauge symmetry) then simply has the status of a clever
calculational trick: it allows the use of a wave function so simple that the
pair-occupation amplitudes $v_j$ can be found with a minimum of effort. The
trick's justification is that the corrections $\delta v_j$'s produced by more
careful approaches usually are small. (The device of using symmetry-breaking
wave-functions purely for the sake of calculational convenience is widespread
in nuclear physics, and lucidly discussed in Ring and Schuck's
book\cite{RingSchuck-80} in a chapter entitled ``Restoration of Broken
Symmetries''.)

The above conclusions imply that the following strategy should suffice for a
{\em qualitative}\/ description (more is not attempted here) of pairing
correlations in isolated ultrasmall grains: although strictly speaking a
fixed-$N$ technique would be appropriate, we shall adopt BCS's grand-canonical
approach throughout, using $u_j, v_j$ as grand-canonical approximations to
$\bar u_j, \bar v_j$.  Quantitatively, this strategy is expected to become
unreliable in the limit of large level spacing $d / \tilde \Delta > 1$
(corresponding to ``strong coupling'' in nuclear applications).  However, the
corrections due to a fixed-$N$ calculation (currently under investigation
applying projection\cite{Braun-98} and exact
diagonalization\cite{Fazio-97} methods), which should become significant in this
regime, are not expected to be more severe than, for example, corrections
arising from a non-equidistant level spectrum, which qualitatively are
insignificant \cite{Smith-96}.

\section{Generalized Variational BCS Approach}
\label{generalBCS}

Since in RBT's experiments $T= 50\mbox{mK} \ll d, \tilde \Delta$, we set
$T=0$.  Our goal in this section is to calculate the discrete eigenenergies of
an isolated, nm-scale metallic grain with pairing correlations, and understand
their evolution in a magnetic field.  To this end, we study the simplest
conceivable pairing model within a generalized variational BCS approach.  The
results will be used in the next section as input into the calculation of the
SET tunneling spectrum of such a grain (see
Fig.~\ref{fig:experimental-spectra} below).

\subsection{The Model}

The only symmetry expected to hold in realistic, irregularly-shaped ultrasmall
grains at zero magnetic field is time-reversal symmetry. We therefore adopt a
single-particle basis of pairs of time-reversed states $|j \pm \rangle$, whose
discrete energies $\varepsilon_j$ are assumed to already incorporate the
effects of impurity scattering and the average of electron-electron
interactions, etc.  As simplest conceivable model describing a pairing
interaction and a Zeeman coupling to a magnetic field, we adopt the following
(reduced) BCS Hamiltonian \cite{vonDelft-96,Braun-97}:
\begin{equation} 
  \label{eq:hamilton}
  \op{H} = \sum_{j, \sigma= \pm} (\varepsilon_j \! - \! \mu \! +\!  \sigma h) 
    c^\dagger_{j\sigma}c_{j\sigma} 
    - \lambda d \sum_{j,j'}
    c^\dagger_{j+}c^\dagger_{j-}c_{j'-}c_{j'+} \; .
\end{equation}
Due to level repulsion the $\varepsilon_j$'s will be approximately uniformly
spaced. For simplicity, we take a completely uniform spectrum with {\em level
  spacing\/} $d$, $\varepsilon_j = j d + \varepsilon_0$. Fluctuations in the
level spacings have been studied with methods of random matrix
theory\cite{Smith-96}, with qualitatively similar results.  For a system with
a total of $N=2m +p$ electrons, where the {\em electron number parity} $p$ is
$0$ for even $N$ and $1$ for odd $N$, we use the label $j=0$ for the first
level whose occupation in the $T=0$ Fermi sea is not 2 but $p$.

The pairing interaction is taken to include only states with $|\varepsilon_j|
< \omega_c$.  Experimental evidence for the sufficiency of neglecting
couplings between non-time-reversed pairs of states, i.e.\ of using only a
{\em reduced}\/ BCS-Hamiltonian, will be given in
section~\ref{sec:time-reversed}.  For convenience we wrote the pair-coupling
constant in Eq.~(\ref{eq:hamilton-1}) as $V=\lambda d$, where $\lambda$ is a
dimensionless parameter. The $d\to 0$ ``bulk gap'' of the model thus is
$\tilde \Delta =2 \omega_c e^{- 1/\lambda}$.

An applied magnetic field will completely penetrate an ultrasmall grain, since
its radius (typically $r \simeq5$nm) is much smaller than the penetration
length of 50 nm for bulk Al. The Zeeman term in Eq.~(\ref{eq:hamilton}), with
$\pm h \equiv \pm \frac12 \mu_Bg H$, models the fact that the measured tunnel
spectra of RBT \cite{Black-96,Ralph-97} (shown in
Fig.~\ref{fig:experimental-spectra} in section~\ref{sec:tunneling-spectra})
evolve approximately linearly as a function of magnetic field, with
$g$-factors between 1.95 and 2 (determined from the differences between
measured slopes of up- and down-moving lines).  Deviations from $g = 2$
probably result from spin-orbit scattering, known to be small but non-zero in
thin Al films \cite{Meservey-94}, but neglected below (where $g=2$ is used).
Furthermore, orbital diamagnetism is also negligible, just as for thin films
in a parallel magnetic field \cite{Meservey-94} but in marked contrast to bulk
samples where it causes the Meissner effect: the grains are so small that even
a $7$T field produces a flux through the grain of only about 5\% of a flux
quantum $\phi_0$, which is too small to significantly affect the orbital
motion of the electrons between subsequent reflections off the grain boundary.
Some larger grains do show slight deviations from $H$-linearity
\cite{Black-96}, which probably reflect the onset of orbital magnetism (which
gives corrections\cite{Bahcall-priv} to the eigenenergies of the order of
$\hbar v_F r^3(H/\phi_0)^2$); however, these effects are much smaller than
Zeeman energies in the grains of present interest, and will be neglected here.
Thus, our model assumes that Pauli paramagnetism due to the Zeeman energy
completely dominates orbital diamagnetism, similarly to the case of thin films
in parallel magnetic fields \cite{Meservey-94}.

\subsection{The Variational Ansatz}

The Zeeman term favours states with non-zero total $z$-component of the total
spin $s=\sum_j s^z_j$ (henceforth simply called ``spin''), so that increasing
$h$ will eventually lead to a series of ground state changes to states with
successively larger spins.  Therefore, we are interested in general in
correlated states with non-zero spin, and in particular in their
eigenenergies.  We calculate these variationally, using the following general
Ansatz for a state $|s,\bbalpha \rangle$ with a definite total spin $s$
(introduced by Soloviev for application in nuclei \cite{Soloviev-61}):
\begin{eqnarray}
  \label{eq:ansatz}
    |s, \bbalpha\rangle = \prod_{j=1}^{2s} c^\dagger_{\alpha(j)+} 
                \prod'_i (u^{(s,\bbalpha)}_i + v^{(s,\bbalpha)}_i
       c^\dagger_{i+}c^\dagger_{i-})\,|\mbox{Vac}\rangle \; .
\end{eqnarray}                                               
The non-zero spin is achieved by placing $2s$ unpaired spin-up electrons in a
set of 2s single particle states, say with labels $j =
\alpha(1),\alpha(2),\cdots,\alpha(2s)$ (see Fig.~\ref{fig:alpha-states}),
while the remaining single-particle pairs of states have BCS-like amplitudes
to be either filled $(v^{(s,\bbalpha)}_i)$ or empty $(u^{(s,\bbalpha)}_i)$,
with $(u^{(s,\bbalpha)}_i)^2+(v^{(s,\bbalpha)}_i)^2=1$.  The prime over
products (and over sums below) indicates exclusion of the singly occupied
states $\alpha(1),\alpha(2),\cdots,\alpha(2s)$ (for which
$u^{(s,\bbalpha)},v^{(s,\bbalpha)}$ are not defined).

A short standard calculation reveals that the constructed wave functions are
orthogonal: $\langle s, \bbalpha | s', \bbalpha'\rangle =
\delta_{ss'}\delta_{\bbalpha\bbalpha'}$. Therefore, the variational parameters
$v_j^{(s,\bbalpha)}$ and $u_j^{(s,\bbalpha)}$ must be found {\em
  independently\/} for each $(s, \bbalpha)$ (hence the superscript). This is
done by minimizing the variational ``eigenenergies''
\begin{eqnarray}
  \label{eq:Esalpha}
\lefteqn{\E_{s,\bbalpha} (h,d) \; \equiv \; 
    \langle s,\bbalpha | H | s,\bbalpha \rangle} \nonumber\\
    &=& 
    -2sh + \sum_{j=1}^{2s}\varepsilon_{\alpha(j)}
    + 2\sum'_j\varepsilon_j(v^{(s,\bbalpha)}_j)^2 - \\
 &{}& - \lambda d\Big(\sum'_{j}u^{(s,\bbalpha)}_jv^{(s,\bbalpha)}_j\Big)^2 
    + \lambda d \sum'_j (v^{(s,\bbalpha)}_j)^4, \nonumber
\end{eqnarray}
which we use to approximate the model's exact eigenenergies
$E_{s,\bbalpha}(h,d)$.  Note that singly-occupied states are excluded from all
primed sums involving $u_j$'s and $v_j$'s. The last term, proportional to
$v^4$, is not extensive and hence neglected in the bulk case where only
effects proportional to the system volume are of interest. Here we retain it,
since in ultrasmall systems it is non-negligible (but not dominant either).

Solving the energy-minimization conditions 
\begin{eqnarray}
  \label{eq:extremal}
  \partial {\cal E}_{s,\bbalpha}/\partial v^{(s,\bbalpha)}_j = 0
\end{eqnarray}
in standard BCS fashion yields 
\begin{equation}
\label{vj}
(v_j^{(s,\bbalpha)})^2 = (1 - \xi_j / [\xi_j^2
+ \Delta_{s,\bbalpha}^2]^{1/2})/2 ,
\end{equation} 
where the ``pairing parameter'' $\Delta_{s,\bbalpha}$ is determined by the
generalized ``gap equation''
\begin{eqnarray}
  \label{eq:gap1}
  \Delta_{s,\bbalpha} & = & \lambda d \sum'_j u_j^{(s,\bbalpha)}
  v_j^{(s,\bbalpha)},\qquad\mbox{or}
\\
  \label{eq:gap}
  \frac1\lambda & = & d\sum'_j \frac1{2 \sqrt{\xi_j^2+\Delta_{s,\bbalpha}^2}},
\end{eqnarray}
and $\xi_j \equiv \varepsilon_j-\mu - \lambda d (v_j^{(s,\bbalpha)})^2$. Note
that we retain the $ \lambda d (v_j^{(s,\bbalpha)})^2$ shift in $\xi_j$,
usually neglected because it simply renormalizes the bare energies, since for
large $d$ it somewhat increases the effective level spacing near
$\varepsilon_F$ (and its neglect turns out to produce a significant upward
shift in the $\E_{s,\bbalpha} (h,d)$'s, which one is trying to minimize).  The
chemical potential $\mu$ is fixed by requiring that
\begin{eqnarray}
  \label{eq:mu}
   2m + p =  \langle s,\bbalpha | \hat N | s,\bbalpha \rangle
  = 2s + 2\sum_j'(v_j^{(s,\bbalpha)})^2 \; .
\end{eqnarray} 
Generally Eqs.~(\ref{vj}), (\ref{eq:gap}) and (\ref{eq:mu}) have to be solved
simultaneously numerically.  In the limit $d/\tilde \Delta \to 0$
(investigated analytically in Appendix~\ref{dto0}), Eq.~(\ref{eq:gap}) reduces
to the standard bulk $T=0$ gap equation.

In contrast to conventional BCS theory, the pairing parameter
$\Delta_{s,\bbalpha}$ can in general not be interpreted as an energy gap and
is {\em not}\/ an observable. It should be viewed simply as a mathematical
auxiliary quantity which was introduced to conveniently solve
Eq.~(\ref{eq:extremal}). However, by parameterizing the variational quantities
$v_j^{(s,\bbalpha)}$ and $u_j^{(s,\bbalpha)}$, $\Delta_{s,\bbalpha}$ does
serve as a measure of the pairing correlations present in a state
$|s,\bbalpha\rangle$, since for vanishing $\Delta_{s,\bbalpha}$ the latter
reduces to an uncorrelated paramagnetic state with spin $s$, namely
\begin{eqnarray}
  \label{eq:param}
  |s,\bbalpha\rangle_0 \equiv \prod_{j=1}^{2s}c_{\alpha(j)+}^\dagger 
                          \prod_{i< 0}'c_{i+}^\dagger c_{i-}^\dagger |0\rangle.
\end{eqnarray}
We shall denote the energy of this uncorrelated state by $\E^0_{s,\bbalpha} =
{}_0\langle s,\bbalpha|H|s,\bbalpha\rangle_0$, and define the ``correlation
energy'' of $|s,\bbalpha\rangle$ as the energy difference ${\cal E}^{corr}_{s,
  \bbalpha} \equiv \E_{s,\bbalpha}-\E^0_{s,\bbalpha}$.

\subsection{Qualitative Discussion}

Before launching into numerical results, let us anticipate by qualitative
arguments what is to be expected:

Firstly, the gap equation for $\Delta_{s,\bbalpha}(d)$ is $h$-{\em
  in}\/dependent.  The reason is that only those $j$-levels contribute in the
gap equation that involve correlated {\em pairs}\/ of states, each of which
have spin 0 and hence no Zeeman energy. Consequently, the $-2sh$-dependence of
$\E_{s,\bbalpha}$ in Eq.~\ref{eq:Esalpha} is simply that of the $2s$ unpaired
electrons.

\begin{figure}
  \begin{center}
    \leavevmode \epsfxsize 7cm
    \epsffile{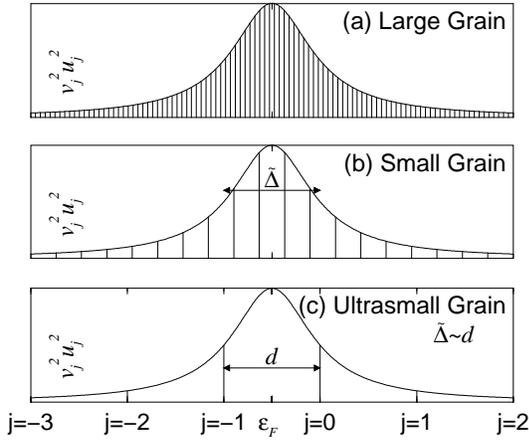}
    \vspace*{5mm} 
    \caption{A cartoon depiction of why ``superconductivity
      breaks down'' when the sample becomes sufficiently small. Each vertical
      line represents a pair of single-particle state $|j \pm \rangle$ with
      energy $\varepsilon_j$, for three different mean level spacings $\ddd$,
      corresponding to (a) a ``large'' grain ($d\ll\tilde\Delta$); (b) a
      ``small'' grain ($\ddd \simeq 0.25 \tilde \Delta$); (c) an
      ``ultrasmall'' grain ($\ddd \simeq \tilde \Delta$).  In all three plots,
      the height of each vertical line equals the function $u_j^2 v_j^2 =
      \frac14 {\tilde \Delta^2 \over \varepsilon_j^2 + \tilde \Delta^2}$ of
      standard bulk BCS theory, illustrating the energy-regime (of range
      $\tilde \Delta$ around $\varepsilon_F$) within which electrons are
      affected by pairing correlations.  Loosely speaking, the number of
      single-electron states $ \tilde \Delta /\ddd$ in this regime corresponds
      to ``the number of Cooper pairs'' of the system.  Evidently, when $\ddd
      / \tilde \Delta \stackrel{\scriptscriptstyle >}{\scriptscriptstyle \sim}
      1$ as in (c), ``the number of Cooper pairs'' becomes less than one and
      it no longer makes sense to call the system ``superconducting''.  }
    \label{fig:v2u2}
  \end{center}
\end{figure}

\begin{figure}
  \begin{center}
    \leavevmode \epsfxsize 7cm
    \epsffile{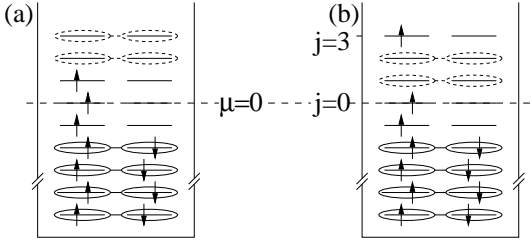}
    \vspace*{5mm} 
    \caption{Two examples of states in the spin-$\frac32$ sector
      of Hilbert space: (a) the ground state $|\frac32\rangle$ and (b) the
      excited state $|\frac32, 2 \rangle$.  The single-particle levels are
      drawn at $h=0$, and we indicated schematically how states are paired
      according to $(u_i + v_i c^\dagger_{i+}c^\dagger_{i-})$ in the BCS-like
      Ans\"atze~(\ref{eq:ansatzs}) and (\ref{eq:ansatzs12}) for
      $|\frac32\rangle$ and $|\frac32, 2 \rangle$, with solid or dashed
      ellipses connecting states that would be completely filled or empty in
      the absence of pairing correlations.}
    \label{fig:alpha-states}
  \end{center}
\end{figure}

Secondly, the discreteness of the sum in the gap equation (\ref{eq:gap}) will
cause $\Delta_{s,\bbalpha}$ to decrease with increasing $d$.  To see this,
inspect Fig.~\ref{fig:v2u2}, in which the height of each vertical line
represents the value of $u_j v_j $ for a time-reversed pair $|j \pm \rangle$.
Figs.~\ref{fig:v2u2}(a) to (c) illustrate that an increase in level spacing
implies a decrease in the number of pairs with significant pair-mixing, i.e.
those within $\tilde \Delta$ of $\varepsilon_F$ which have non-zero $u_j v_j$.
This number can roughly speaking be called the ``number of Cooper pairs'' of
the system.  Since for $d \gg \tilde \Delta$ {\em no}\/ pairs lie in the
correlated regime $|\varepsilon_j - \varepsilon_F| < \tilde \Delta$ where
pair-mixing occurs, $\Delta_{s , \bbalpha}$ will be zero in this limit, so
that in general $\Delta_{s , \bbalpha} (d)$ will be a decreasing function of
$d$, dropping to zero at about $d \simeq \tilde \Delta$. Physically speaking,
this happens since with increasing $d$ the increasing kinetic energy cost of
pair-mixing (which shifts pair-occupation amplitude from below to above
$\varepsilon_F$) causes the correlations to weaken, becoming negligible for
large enough $d$.

Thirdly, the $(s,\bbalpha)$-dependent restriction on the primed sum in the gap
equation implies that $\Delta_{s , \bbalpha} (d)$ at fixed $d$ will decrease
with increasing $s$: larger $s$ means more unpaired electrons, more terms
missing from the primed sum, less correlated pairs and hence smaller
$\Delta_{s , \bbalpha}$.  The physics behind this has been called the {\em
  blocking effect\/}\cite{Soloviev-61} in nuclear physics: Singly-occupied
states cannot take part in the pair-scattering caused by the BCS-like
interaction (\ref{eq:hamilton}) and hence decrease the phase space for pair
scattering, as explained in section \ref{p:blocking}.  (Their absence in the
primed sum simply reflects this fact.)  The blocking effect becomes stronger
with increasing $d$, since then the relative weight of each term missing in
the primed sum increases.  It also is stronger the closer the blocked state
lies to $\varepsilon_F$, since the excluded $u_j^{(s,\bbalpha)}
v_j^{(s,\bbalpha)}$ contribution to the primed sum is largest near
$\varepsilon_F$, as is evident from Fig.~\ref{fig:v2u2}.  On the other hand,
an unpaired electron will have almost no blocking effect if $|\varepsilon_j -
\varepsilon_F| \gg \tilde \Delta$, since $u_j^{(s,\bbalpha)}
v_j^{(s,\bbalpha)}$ vanishes there anyway.

Finally, note that the $(s,\bbalpha)$-dependence of $\Delta_{s,\bbalpha}$ for $d
\simeq \tilde \Delta$ illustrates why in this regime a conventional mean-field
treatment is no longer sufficient: the system cannot be characterized by a
single pairing parameter, since the amount of pairing correlations vary from
state to state, each of which is characterized by its own pairing parameter.

\subsection{General Numerical Solution}
\label{generalnumerics}

It is possible to solve the modified gap equation analytically in two limits,
$d\ll\tilde\Delta$ and $d\gg\Delta_s$ (see Appendix~\ref{analytics}), but
generally the gap equation and (\ref{eq:mu}) have to be solved numerically.
In doing so, some assumptions are necessary about parameter values (though
using slightly different values would not change the results qualitatively).
We measure all energies in units of the bulk gap $\tilde \Delta =2 \omega_c
e^{- 1/\lambda}$ of the model. However, its experimental value differs from
that of a truly bulk system, since it is known from work with Al thin films
\cite{Strongin-70,Garland-68} that the effective dimensionless
pairing-interaction strength $\lambda$ is larger in Al samples of reduced
dimensionality than in truly bulk, three-dimensional systems. (Though true for
Al, this is not a universal property of small samples, though --- for Nb,
$\tilde \Delta$ is larger in the bulk than in thin films\cite{Ralph-priv}.)
Since thin films in a parallel magnetic field are analogous in many ways to
ultrasmall grains (in particular regarding the dominance of Pauli
paramagnetism over orbital diamagnetism), we shall assume that the effective
coupling constant $\lambda$ is the same in both.  Adopting, therefore, the
value $\tilde\Delta=0.38$meV found for thin Al films in
Ref.~\onlinecite{Meservey-70}, and taking the cut-off to be the Debye
frequency $\omega_c=33$meV of Al, we use $\lambda= [\ln(2 \omega_c/ \tilde
\Delta)]^{-1}=0.224$ for the dimensionless pairing-interaction strength.
Furthermore, we smeared the cutoff of the BCS interaction over two
single-electron levels, to ensure that discontinuities do not occur in
$d$-dependent quantities each time the energy $|\varepsilon_j = d j +
\varepsilon_0|$ of some large-$|j|$ level moves beyond the cut-off $\omega_c$
as $d$ is increased.

Solving Eqs.~(\ref{vj}), (\ref{eq:gap}) and (\ref{eq:mu}) is a straightforward
numerical exercise which we performed, for the sake of ``numerical
consistency'', without further approximations.  (Since some minor
approximations were made in Ref.~\onlinecite{Braun-97}, e.g.\ dropping the
$\lambda d v_j^2$ term in $\xi_j$, and slightly different parameter-values
were used, the numerical results there sometimes differ slightly from the
present ones; see, e.g.\ Fig.~\ref{fig:pairing-parameter}.)  It should be
understood, though, that only qualitative significance can be attached to our
numerical results, since our model is very crude: it neglects, for instance,
fluctuations in level spacing and in pair-coupling constants, and we do not
carry out a fixed-$N$ projection, all of which presumable would somewhat
influence the results quantitatively.

\subsubsection{Spin-$s$ Ground States}

\begin{figure}
  \begin{center}
    \leavevmode \epsfxsize 8cm
    \epsffile{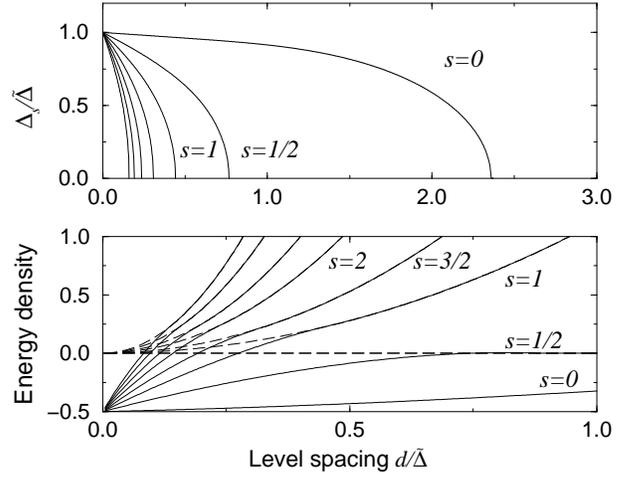}
    \vspace*{5mm}  
    \caption{Properties of spin-$s$ ground states
      $|s\rangle$ \mbox{[}compare Eq.~(\protect\ref{eq:ansatzs})\mbox{]}: (a)
      The pairing parameters $\Delta_s(d)/ \tilde \Delta$ for some spin-$s$
      ground states $|s\rangle$, as a functions of $d/\tilde \Delta$.  The
      critical level spacings $d_{c,s}$ at which $\Delta_s (d_{c,s}) =0$ are
      found to be 2.36, 0.77, 0.44, 0.31, \dots for $s = 0, 1/2, 1, 3/2,
      \dots$, respectively.  (b) The energy densities $({\cal E}_s - {\cal
        E}^0_{p/2})d / \tilde \Delta^2$ (solid lines), plotted as functions of
      $d/\tilde \Delta$ for $h=0$, of some pair-correlated spin-$s$ ground
      states $|s \rangle$ relative to the uncorrelated spin-$p/2$ Fermi sea
      $|p/2\rangle_0$, and for comparision the relative energy densities
      $({\cal E}^0_s - {\cal E}^0_{p/2}) d / \tilde \Delta^2$ (dashed lines)
      of the corresponding uncorrelated paramagnetic states $|s\rangle_0$
      (obtained from $|s\rangle$ by setting $\Delta_s = 0$).  We call the
      plotted quantities energy densities since the normalization factor
      $d/\tilde \Delta^2$ contains $d \sim \mbox{Vol}^{-1}$. The solid and
      dashed spin-$s$ lines meet at the critical level spacing $d_{c,s}$,
      above which no pairing correlations survive (so that the relative energy
      densities equal $(s^2 - p/4+(s-p/2)\lambda) d^2/\tilde \Delta^2$
      there).}
    \label{fig:pairing-parameter} \vspace*{5mm}
  \end{center}
\end{figure}

In a given spin-$s$ sector of Hilbert space (with $p = 2s\,\mbox{mod}\,2$),
let $s \rangle $ be the variational state with the lowest energy, i.e.\ the
``variational spin-$s$ ground state''. It is obtained by placing the $2s$
unpaired electrons as close as possible to $\varepsilon_F$
[Fig.~\ref{fig:alpha-states}(a)], because this minimizes the kinetic energy
cost of having more spin ups than downs:
\begin{eqnarray}
  \label{eq:ansatzs}
    |s\rangle = \!\!\!\!
    \prod_{j=-s + p/2}^{s-1+p/2} \!\!\!\! c^\dagger_{j +} 
                \prod'_i (u^{s}_i + v^{s}_i
       c^\dagger_{i+}c^\dagger_{i-})\,|\mbox{Vac}\rangle \; .
\end{eqnarray}       
(The particular choice of $\bbalpha$ in the general Ansatz (\ref{eq:ansatz})
to which $|s\rangle$ corresponds is $\alpha(n) = n-[s]-1$ for $n=1\ldots 2s$,
where $[s]$ is the largest integer $\le s$.)  The numerical results for the
corresponding pairing parameters $\Delta_s (d)$, shown in
Fig.~\ref{fig:pairing-parameter}(a) for some several small $s$, confirm the
properties anticipated in the previous subsection's qualitative discussion:

Firstly, each $\Delta_s$ decreases with $d$, vanishing at a critical level
spacing $ d_{c,s}$ beyond which no pair-mixing correlations exist in this
level of approximation.  In Appendix~\ref{sec:neardc} it is shown that near
$d_{c,s}$, $\Delta_s(d)$ has the standard mean-field form
$\sqrt{1-d/d_{c,s}}$; this was to be expected, since the variational approach
to finding $|s \rangle$ is equivalent to doing standard mean-field theory
within the spin-$s$ sector of Hilbert space.  (Note that one should not attach
too much significance to the precise numerical values of the $d_{c,s}$
reported in Fig.~\ref{fig:pairing-parameter}, since they depend sensitively on
model assumptions: for example, the values for $d_{c,0}$ and $d_{c,1/2}$
differ somewhat from those reported in
Refs.~\onlinecite{vonDelft-96,Braun-97}, due to their use of a slightly
different $\lambda$ and minor numerical approximations not used here, as
mentioned above. Moreover, Smith and Ambegaokar\cite{Smith-96} showed that the
precise distribution of levels used influences $d_{c,s}$ significantly.)

Secondly, $\Delta_s$ decreases rapidly with increasing $s$ at fixed
$d$ (and $d_{c,s} < d_{c,s'}$ if $s>s'$), illustrating the blocking
effect. This result, which is expected to be independent of model
details, is a generalization of the parity effect discussed by von
Delft {\em et al.}\cite{vonDelft-96}.  (They studied only ground state
pairing correlations and found that these are weaker in odd $(s=1/2)$
grains than in even $(s=0)$ grains, $\Delta_{odd} = \Delta_{1/2} <
\Delta_{even}=\Delta_0$.)  The blocking effect is most dramatic in the
regime $d/\tilde\Delta\in[0.77, 2.36]$ in which $\Delta_{0} \neq 0 $
but $\Delta_{s \neq 0 } = 0$. This is a regime of ``minimal
superconductivity''\cite{Braun-97}, in the sense that all pairing
correlations that still exist in the even ground state (since
$\Delta_0 \neq 0$) are completely destroyed by the addition of a
single electron or the flipping of a single spin (since
$\Delta_{s \neq 0} = 0$).

Fig.~\ref{fig:pairing-parameter}(b) shows the eigenenergies ${\cal E}_{s}$
(solid lines) of $|s \rangle$ and the energies ${\cal E}^0_{s}$ (dotted lines)
of the corresponding uncorrelated paramagnetic states
\begin{equation}
  \label{eq:ansatzs00}
    |s\rangle_0 = \!\!\!\!
       \prod_{j=-s + p/2}^{s-1+p/2} \!\!\!\! c^\dagger_{j +} 
                \prod_{i< -s + p/2} c^\dagger_{i+}c^\dagger_{i-}\,
                |\mbox{Vac}\rangle .
\end{equation}
The solid and dashed spin-$s$ lines meet at the critical level spacing
$d_{c,s}$, above which no pairing correlations survive.

\subsubsection{Spin-$s$ Excited States}
\label{excited}

Among all possible excited states with definite $s$, we consider here only
those created from $|s\rangle$ by exiting one electron from the topmost
occupied level $ s-1+p/2$ of $s\rangle$ to some higher level $j + s-1+p/2$:
\begin{eqnarray}
  \label{eq:ansatzs12}
    |s,j\rangle &=&  
    c^\dagger_{(j + s-1+p/2) +} \!\!
    \prod_{\bar j=-s + p/2}^{s-2+p/2} \!\!\!\!
    c^\dagger_{\bar j +}  \\
    & & 
            \quad \times    \prod'_i (u^{s}_i + v^{s}_i
       c^\dagger_{i+}c^\dagger_{i-})\,|\mbox{Vac}\rangle \; .
\end{eqnarray}       
(This reduces to $|s\rangle$ if $j=0$; the particular choice of $\bbalpha$ in
Ansatz (\ref{eq:ansatz}) to which $|s,j \rangle$ corresponds is $\alpha(n) =
n-[s]-1$ for $n=1\ldots 2s-1$ and $\alpha(2s) = [s]-1+j$.)

\begin{figure}
  \begin{center}
    \leavevmode \epsfxsize 7cm
    \epsffile{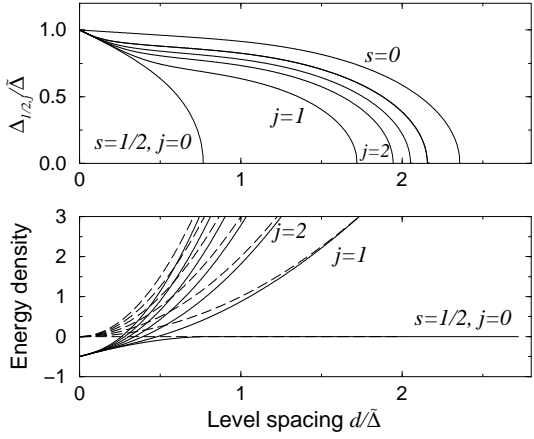}
    \vspace*{5mm} \caption{Properties of excited spin-$\frac12$ states
      $|\frac12, j\rangle$ [compare Eq.~(\ref{eq:ansatzs12})]: (a) The pairing
      parameter $\Delta_{1/2, j}$ for some spin-$\frac12$ states
      $|\frac12,j\rangle$ $(j = 0, \dots, 4)$, together with $\Delta_0$ of the
      spin-$0$ ground state $|0 \rangle$ (the outermost curve).  The larger
      $j$, the closer $\Delta_{1/2, j}$ approaches the spin-0 value
      $\Delta_0$. (b) The relative energy densities $({\cal E}_{1/2,j} - {\cal
        E}^0_{1/2,0}) d / \tilde \Delta^2$ (solid lines) of
      $|\frac12,j\rangle$ relative to $|\frac12,0\rangle_0 = |\frac12\rangle_0$,
      and for comparison the relative energy densities $({\cal E}^0_{1/2,j} -
      {\cal E}^0_{1/2,0}) d / \tilde \Delta^2$ (dashed lines) of the
      corresponding uncorrelated state $|\frac12,j\rangle_0$.  For excited
      states the solid and dashed lines meet at a larger $d$ than for the
      ground state, i.e. in excited states pairing correlations survive down
      to smaller grain sizes than in the corresponding ground state.}
    \label{fig:excited-states} \vspace*{5mm}
  \end{center}
\end{figure}

Interestingly, one finds that the larger $j$, the longer the pairing
correlations survive with increasing $d$.  This is illustrated by the simple
example $s=1/2$: Fig.~\ref{fig:excited-states}(a) shows that the critical
spacings $d_{c,1/2,j} $ (at which the pairing parameters $\Delta_{1/2,j} (d) $
vanish) increase with $j$, approaching the value $d_{c,0}$ of the spin-0 case
as $j \to \infty$.  This result is reflected in the excitation energies of
Fig.~\ref{fig:excited-states}(b): the excited states of a the spin-1/2 sector
have non-zero correlation energies (difference between solid and dashed lines)
at $d$-values for which the spin-$1/2$ ground state correlation energy of
Fig.~\ref{fig:pairing-parameter}(b) is already zero.  The intuitive reason why
more-highly-excited states have more pairing correlations than the
corresponding spin-$1/2$ ground state $|1/2\rangle$ is of course quite simple:
The larger $j$, i.e.\ the further the unpaired electron sits from the Fermi
surface where pairing correlations are strongest, the less it disrupts
pair-mixing (since $u_j v_j$ becomes very small for large $j$, see
Fig.~\ref{fig:v2u2}).  In fact, for very large $j$, the state $|\frac12,j
\rangle$ will have the just about same amount of pairing correlations as the
even ground state $|0 \rangle$ ($\Delta_{1/2,j} \simeq \Delta_0$), since the
unpaired electron sits so far from $\varepsilon_F$ that the pairing
correlations are effectively identical to those of $|0 \rangle$.

Similar effects are seen for excited states in other spin sectors
$s\neq\frac12$. The higher the excitation, the larger the pairing parameter
$\Delta_{s,\bbalpha}$. Nevertheless the energy of the excited states is always
higher than that of the corresponding spin-$s$ ground state, since the
kinetic-energy cost of having an unpaired electron far from $\varepsilon_F$
can be shown to always outweigh the interaction-energy gain due to having less
blocking and hence a larger $\Delta_{s,\bbalpha}$.

\subsection{Magnetic Field Behaviour}
\label{magfield}

In a magnetic field, the Zeeman energy favors states with non-zero spin.
However, since such states have smaller correlation energy due to the blocking
effect a competition arises between Zeeman energy and correlation energy.  The
manifestations of the blocking effect can thus be probed by turning on a
magnetic field; if it becomes large enough to enforce a large spin, excessive
blocking will destroy all pairing correlations.

The situation is analogous to ultra-thin films in a parallel magnetic
field,\cite{Meservey-94} where orbital diamagnetism is negligible for
geometrical reasons and superconductivity is destroyed at sufficiently large
$h$ by Pauli paramagnetism.  This occurs via a first order transition to a
paramagnetic state, as predicted by Clogston and Chandrasekhar (CC)
\cite{Chandrasekhar-62,Clogston-62} by the following argument (for bulk
systems): A pure Pauli paramagnet has ground state energy $-h^2{\cal
  N}(\varepsilon_F)$ and spin $s = h {\cal N}(\varepsilon_F)$ (since it
chooses its spin such that the sum of the kinetic and Zeeman energies at spin
$s$, $s^2 {\cal N}(\varepsilon_F) - 2 h s$, is minimized).  When this energy
drops below the bulk correlation energy $-\frac12\tilde\Delta^2{\cal
  N}(\varepsilon_F)$ of the superconducting ground state, which happens at the
critical field $h_{CC} =\tilde\Delta/\sqrt{2}$, a transition will occur from
the superconducting to the paramagnetic ground state. The transition is
first-order, since the change in spin, from 0 to $ s_{CC} = h_{CC} {\cal
  N}(\varepsilon_F) = \tilde \Delta / (d \sqrt 2)$, is macroscopically large
(${\cal N}(\varepsilon_F) = 1/d \simeq \mbox{Vol}$).  In tunneling experiments
into ultra-thin (5nm) Al films ($\tilde\Delta = 0.38$meV and $H_{CC} = 4.7$T)
this transition has been observed\cite{Meservey-70} as a jump in the tunneling
threshold (from $\tilde\Delta-h_{CC}$ to zero) at $h_{CC}$.

In isolated ultrasmall grains, the above picture of the transition needs to be
rethought in two respects due to the discreteness of the electronic spectrum:
Firstly, the spin must be treated as a discrete (instead of continuous)
variable, whose changes with increasing $h$ can only take on
(parity-conserving) integer values.  Secondly, one needs to consider more
carefully the possibility of $h$-induced transitions to non-zero spin states
that are still {\em pair-correlated} (instead of being purely paramagnetic),
such as the variational states $ |s, \bbalpha\rangle $ discussed above.  (In
the bulk case, it is obvious that such states play no role: the lowest
pair-correlated state with non-zero spin obtainable from the ground state by
spin flips is a two-quasi-particle state, costing energy $2\tilde \Delta -
2h$; when $h$ is increased from 0, the paramagnetic transition at $h_{CC} =
\tilde \Delta/\sqrt 2$ thus occurs before a transition to this state, which
would require $h = \tilde \Delta$, can occur.)

Within our variational approach, the effect of increasing $h$ from 0 can be
analyzed as follows: At given $d$ and $h$, the grain's ground state is the
lowest-energy state among all possible spin-$s$ ground states $|s \rangle$
having the correct parity $2s\,\mbox{mod}\,2 = p$.  Since $\E_{s}(h,d) =
\E_{s}(0,d) - 2hs$, level crossings occur with increasing $h$, with $ \E_{s'}$
dropping below $\E_s$ at the {\em level crossing field}\/
\begin{eqnarray}
  \label{eq:hcrit}
  h_{s,s'} (d) = \frac{\E_{s'}(0,d)-\E_s(0,d)}{2(s'-s)} \; . 
\end{eqnarray}
Therefore, as $h$ is slowly turned on from zero with initial ground state
$|s_0 = p/2\rangle$, a cascade of successive ground-state changes (GSC) to new
ground states $|s_1 \rangle$, $|s_2 \rangle$, \dots will occur at the fields
$h_{s_0, s_1}$, $h_{s_1, s_2}$, \dots We denote this cascade by $(s_0,s_1);
(s_1,s_2);\ldots$, and for each of its ground state changes  the
corresponding level-crossing fields $h_{s,s'} (d)$ is shown in 
 Fig.~\ref{fig:h-crit}. 
Generalizing CC's critical
field to non-zero $d$, we denote the (parity-dependent) field at which the
{\em first}\/ transition $(s_0, s_1)$ occurs by $h_{CC} (d,p) \equiv h_{s_0,
  s_1} (d) $, which simply is the lower envelope of the level-crossing fields
$h_{s_0, s_1}$ in Fig.~\ref{fig:h-crit}.  In the limit $d\to0$ we find
numerically that it correctly reduces to the Clogston-Chandrasekhar value
$h_{CC}(0,p) = \tilde\Delta/\sqrt{2}$.

In general, the order in which the GSCs occur with increasing $h$ depends
sensitively on $d$ and an infinite number of distinct regimes (cascades) I,
II, III, \dots can be distinguished: Starting at large $d$ we find the typical
normal behaviour $(0,1); (1,2); (2,3); \ldots$ for even grains and
$(\frac12,\frac32); (\frac32,\frac52); \ldots$ for odd grains, with $h_{0,1} <
$ (or $>$) $ h_{\frac12,\frac32}$ in regimes I (or II).  In regimes III and IV
of somewhat smaller $d$, the order of GSCs is $(0,2); (2,3); \ldots$ and
$(\frac12,\frac32); (\frac32,\frac52); \ldots$, etc, i.e.\ the spin $s_1$
attained after the first GSC $(s_0,s_1)$ has increased to 2 in the even case.
This illustrates a general trend: the spin $ s_1 (d)$ after the first
transition increases with decreasing $d$ and becomes macroscopically large in
the $d\to0$ limit, where $s_1 = h_{CC} / d = \tilde \Delta / (d \sqrt 2)$, as
explained in recounting CC's argument above.

Furthermore, it turns out that $\Delta_{s_1} (d)= 0$ for {\em all}\/ $d$,
implying that after the first GSC the new ground state $|s_1\rangle$ is {\em
  always}\/ (not only in CC's bulk limit) an uncorrelated, purely paramagnetic
state.  In this regard, CC's picture of the transition remains valid
throughout as $d$ is increased: at $h_{CC} (d,p)$, a transition occurs from
the superconducting ground state to a paramagnetic, uncorrelated state $|s_1
\rangle_0$, the transition being first-order in the sense that $\Delta_{s_1}
(d)= 0$; however, the first-order transition is ``softened'' with increasing
$d$, in the sense that the size of the spin change, $s_1 - s_0,$ decreases
from being macroscopically large in the bulk to being equal 1 at $d \gg \tilde
\Delta$ (regimes I and II).

\subsection{Deficiencies of the Variational Ansatz}

Though the variational method we used to calculate the systems
``eigenenergies'' is expected to yield qualitatively correct results, it does
have some deficiencies:

Firstly, a variational approach by construction only gives an upper bound on
the exact eigenenergies $E_{s, \bbalpha}$. The variational energies ${\cal
  E}_{s, \bbalpha}$ could be lowered further by choosing better trial wave
functions that sample larger parts of a given spin-$s$ Hilbert space, i.e.\ by
including ``fluctuations'' about the chosen states.

Secondly, the abrupt vanishing of the pairing parameters $\Delta_{s, \bbalpha}
(d) \simeq \sqrt{1-d/d_{c,s}}$ at a critical level spacing $d_{s, \bbalpha}$
[see Appendix~\ref{sec:neardc} and Fig.~\ref{fig:pairing-parameter}(a))] is
unphysical: in a finite system, any non-zero pair-interaction constant will
always induce a non-zero amount of pairing correlations, i.e. the canonical
$\bar \Delta'_s (d)$ of Eq.~(\ref{eq:v2u2}) will always be non-zero, though it
could become arbitrarily small for sufficiently large $d$. (This statement is
analogous to stating that ``in a finite system no abrupt phase transition
between a zero and non-zero order parameter occurs.'').  The abrupt,
mean-field-like vanishing of $\Delta_{s, \bbalpha} (d)$ is of course an
artefact, that occurs since the grand-canonical variational Ansatz is
equivalent (at least for the spin-$s$ ground states $|s \rangle$) to doing
mean-field theory in a fixed-$s$ Hilbert space.

Thirdly, the variational states of course are not $\hat N$-eigenstates (though
they do have definite parity), and Eq.~(\ref{eq:mu}) only fixes the {\em
  mean\/} electron number. Our reasons for nevertheless adopting them to
describe an isolated grain were given in section~\ref{sec:fixedN}: a large
body of experience in nuclear physics showed that fixed-$N$ projections
generally produce only minor corrections to the grand-canonical BCS results.
Nonetheless, note that we expect a fixed-$N$ projection (currently under
investigation \cite{Braun-98}) to somewhat ameliorate the first two
of the above-mentioned deficiencies of the variational approach: projection
after variation of $|s \rangle$ to fixed $N$ will lower the energy ${\cal
  E}_s$ a bit, and presumably projection before variation will in addition
result in a canonical pairing parameter $\bar \Delta_s (d)$ that decays
smoothly with increasing $d$ from finite to arbitrarily small but non-zero
values.  Note, though, that this is not expected to change the eigenenergies
very much, since the correlation energies rapidly approach zero anyway when
the correlations become weak. In other words, we expect the variational scheme
for calculating eigenenergies to break down only when $\Delta_s$ becomes so
small that it has no experimental relevance any more (to check this in detail,
strictly canonical calculations are needed\cite{Braun-98,Fazio-97}).

\section{Observable Quantities} 
\label{observables}

In this section, we consider the grain coupled to leads as in RBT's SET
experiments. After explaining what kind of information can and can not be
extracted from their data, we turn to the calculation of observable
quantities: (a) We calculate theoretical tunneling spectra and compare these
to RBT's measurements; (b) address the question of the observability of
various parity effects, proposing to search for one involving the
pair-breaking energy; and (c) explain how RBT's data give direct evidence for
the dominance of time-reversed pairing, at least for small fields.

\subsection{Experimental Details}
\label{expdetails}

In RBT's experiments \cite{Ralph-95,Black-96,Ralph-97}, an ultrasmall grain
was used as central island in a SET: it was connected via tunnel barriers to
external leads and capacitively coupled to a gate, and its electronic spectrum
determined by measuring the tunnel current through the grain as a function of
transport voltage ($V$), gate voltage $(V_g)$ and magnetic field ($H=h/\mu_B$,
with $\mu_B = 0.0571$meV/T) at a fixed temperature of 50mK.

The particular grain (Ref.~\onlinecite{Ralph-97}, Figs.~1(b),2,3) with which
we shall compare our theory had the following parameters: Its radius was
estimated as $r \simeq 4.5$nm by assuming the grain to be hemispherical,
implying a volume $\simeq (5.7\mbox{nm})^3$ and a total number of conduction
electrons $N$ of about $3 \times 10^4$.  The crude order-of-magnitude
free-electron estimate $d= 2 \pi^2 \hbar^2 / (m k_F \mbox{Vol})$ for the mean
level spacing near $\varepsilon_F$ yields $d \simeq 0.45$meV.  The SET had
lead-to-grain capacitances $C_1 = 3.5$aF, $C_2=9.4$aF, gate-to-grain
capacitance $C_g=0.09$aF and charging energy $E_C = e^2/2C_{\rm total} =
46$meV.  The tunnel current is on the order of $10^{-10}$A, implying an
average time of $2\cdot10^{-9}$sec between subsequent tunneling processes.

Since the charging energy $E_C$ was very much larger than all other energy
scales, such as the bulk gap ($\tilde\Delta \simeq 0.38$meV), typical values
of the transport voltage ($V {\stackrel{\scriptscriptstyle
    <}{\scriptscriptstyle \sim}} 1$mV) and the temperature, fluctuations in
electron number on the grain are strongly suppressed, so that coherent
superpositions between states with different $N$ need not be considered. The
energy-balance condition that determines through which eigenstates of the
grain electrons can tunnel for given values of transport- and gate voltage
thus involve differences between the eigenenergies of a grain with {\em fixed
  particle number} $N$ or $N\pm1$,
\begin{equation}
\label{energydiff}
\Delta E_{if} \equiv 
(E^{N}_{f} + E_C^N) - (E^{N \pm 1}_{i} + E_C^{N\pm1}) \; , 
\end{equation}
corresponding to the energy cost needed for some rate-limiting electron
tunneling process $|i\rangle_{N\pm1} \to |f\rangle_N$ off or onto the grain.
Here $|f\rangle_N$ denotes a discrete eigenstate of the $N$-electron grain
with eigenenergy $E^N_{f} + E_C^N$. Following the ``orthodox model'' of SET
charging, we take $E_C^N$, the grain's electrostatic energy (relative to a
neutral grain with $N_0$ electrons) as $E_C^N = E_C (N - N_0 - Q_g/e)^2$,
where $Q_g = C_g V_g + const$ is the gate charge, and assume the
Coulomb-interaction to be screened sufficiently well that its sole effect is
to shift {\em all}\/ fixed-$N$ eigenstates by the same constant amount
$E^N_{C}$.  (The latter assumption is somewhat precarious: it becomes worse
with decreasing grain size, and was shown to break down in grains half the
present size.\cite{Agam-97})

RBT were able to extract the energy differences $\Delta E_{if}$ from their
data: the differential conductance $dI/dV$ as function of $V$ at fixed $V_g$
has a peak whenever $eV$ times a known capacitance ratio is equal to one of
the $\Delta E_{if}$'s, at which point another channel for carrying tunneling
current through the grain opens up (the inclusion of the capacitance ratio
takes into account that the voltage drop across each of the two tunnel
junctions can be different if their capacitances are not
identical\cite{Ralph-97}).  Plotting the position of each conductance peak as
function of $h$ gives the so-called experimental tunneling spectrum shown in
Fig.~\ref{fig:experimental-spectra}, in which each line reflects the
$H$-dependence of one of the energy differences $\Delta E (h)$.

It is important to note that the experimental threshold energy at $h=0$ for
the lowest-energy tunneling process ($y$-intercept of the lowest line, the
so-called ``tunneling threshold'') yields no significant information, since it
depends on the grain's change in overall charging energy due to tunneling,
\begin{equation}
\delta E_C = 
E_C^{N } - E_C^{N \pm 1 } = E_C [Q_g/e  - (N -N_0 \pm 
{\textstyle \frac12})],
\end{equation}
which depends (via $Q_g$) in an imprecisely-known way on the adjustable gate
voltage $V_g$. This $V_g$-dependence can usually (e.g.\ in SETs with much
smaller charging energies than here) be quantified precisely by studying the
Coulomb oscillations that occur as function of $V_g$ at fixed $V$.
Unfortunately, in the present case a complication arises\cite{Ralph-priv} due
to the smallness of the gate capacitance: to sweep $Q_g$ through one period of
$2e$, the gate voltage $V_{g}$ must be swept through a range so large ($2e/C_g
\simeq 3.5$V) that during the sweep small ``rigid'' shifts of the entire
tunneling spectrum occur at random values of $V_g$. They presumably are due to
single-electron changes in the charge contained in other metal grains in the
neighborhood of the grain of interest; these changes produce sudden shifts in
the electrostatic potential of the grain, and thus spoil the exact
$2e$-periodicity that would otherwise have been expected for
the spectra.

In contrast to the threshold energy, however, the {\em separations between
  lines}\/,
\begin{equation}
\label{lineseparation}
  \Delta E_{if'} - \Delta E_{if} = E_{f'}^N - E_f^N \; ,
\end{equation}
{\em are} independent of gate voltage and hence known absolutely; they simply
correspond to the differences between eigenenergies of a {\em fixed}\/-$N$
grain, i.e.\ give its {\em fixed}\/-$N$ excitation spectrum, and these are the
quantities that we shall focus on calculating below.

The most notable feature of RBT's measured tunneling spectra is the presence
(absence) of a clear spectroscopic gap $2 \Omega_e \! > \! d$ between the
lowest two lines of the odd-to-even (even-to-odd) measured spectra in
Fig.~\ref{fig:experimental-spectra}(a,b).  This reveals the presence of
pairing correlations: in even grains, all excited states involve at least two
BCS quasi-particles and hence lie significantly above the ground state,
whereas odd grains {\em always}\/ have at least one quasi-particle and
excitations need not overcome an extra gap.

Since the $\Delta E_{if}$'s in Eq.~(\ref{energydiff}) are constructed from
{\em fixed}\/-$N$ and {\em fixed}\/-$N\pm1$ eigenenergies, we shall
approximate these using the variational energies ${\cal E}_{s \bbalpha}$
discussed in previous sections for a completely {\em isolated}\/ grain. (We
thereby make the implicit assumption that the grain's coupling to the leads is
sufficiently weak that this does not affect its eigenenergies,
i.e.\ that the leads act as ``ideal'' probes of the grain.)  The ${\cal
  E}_{s \bbalpha}$ will be used as starting point to discuss various
observable quantities; in particular, we shall make contact with RBT's
experimental results by constructing the theoretical tunnel spectrum (as
function of $h$ and $d$) predicted by our model.

\subsection{The Tunneling Spectrum in an Magnetic Field}
\label{sec:tunneling-spectra}

The kind of tunneling spectrum that results depends in a distinct way
on the specific choice of level spacing $d$ and final-state parity $p$
(i.e.\ the parity of the grain after the rate-limiting tunneling
process has occured).  To calculate the spectrum for given $d$ and
$p$, we proceed as follows below: we first analyze at each magnetic
field $h$ which tunneling processes $|i \rangle_{N\pm1} \to |f
\rangle_{N}$ are possible, then calculate the corresponding energy
costs $\Delta E_{if} (h)$ of Eq.~(\ref{energydiff}) and plot $\Delta
E_{if} (h) - \Delta E_{min} (0)$ as functions of $h$ for various
combinations of $i,f$, each of which gives a line in the spectrum.
We subtract $\Delta E_{min} (0)$, the $h=0$ threshold energy
cost for the lowest-lying transition, since in experiment it depends
on $V_g$ and hence yields no significant information, as explained
above.

Fig.~\ref{fig:spectra} shows four typical examples of such theoretical
tunneling spectra, with some lines labeled by the corresponding $|i \rangle
\to |f \rangle$ transition.

\begin{figure}
  \begin{center}
    \leavevmode \epsfxsize 8cm
    \epsffile{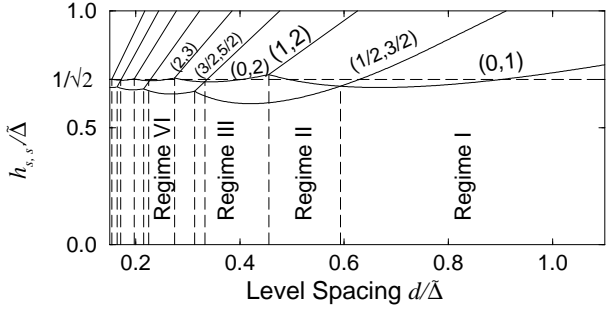}
    \vspace*{5mm} \caption{The level-crossing fields $h_{s,s'} (d)/
      \tilde \Delta$ [see Eq.~(\ref{eq:hcrit})] for the cascade of
      ground state changes (GSCs) $(s_0,s_1)$; $(s_1,s_2)$; $\ldots$ that
      occurs as $h$ increases from 0 at given $d$. Some lines are
      labeled by the associated GSC $(s,s')$ (where ${\cal E}_{s'}$
      drops below ${\cal E}_{s}$ as $h$ increases past $h_{s,s'}$).
      (Level crossing fields {\em not}\/ associated with a GSC are not
      shown.) The order in which GSCs can occur within a cascade
      (i.e.\ the order of $h_{s,s'}$ lines encountered when moving
      vertically upward in the figure) depends sensitively on $d$, and
      an infinite number of distinct regimes (cascades) I, II, III,
      \dots can be distinguished.  }
    \label{fig:h-crit} \vspace*{5mm}
  \end{center}
\end{figure}

When taking the data for Fig.~\ref{fig:experimental-spectra}, RBT took care to
adjust the gate voltage $V_g$ such as to minimize non-equilibrium effects,
which we shall therefore neglect.  For given $h$, we thus consider only those
tunneling processes for which the initial state $|i \rangle$ corresponds to
the grain's ground state $|s_i \rangle$ at that $h$ (and $d$,$p$), whose spin
$s_i$ can be inferred from Fig.~\ref{fig:h-crit}.  Since the grain's large
charging energy ensures that only one electron can tunnel at a time, the set
 $\{ |f\rangle \}$ of possible final states satisfies the ``spin selection
rule'' $|s_f - s_i| = \frac12$ and includes, 
besides the spin-$s_f$ {\em ground}\/ state $|s_f\rangle$, also 
excited  spin-$s_f$ states.

Whenever $h$ passes through one of the level-crossing fields $h_{s_i, s_{i'}}$
of (\ref{eq:hcrit}), the grain experiences a ground state change $(s_i,
s_{i'})$. After this GSC, $|s_{i'} \rangle$ is the new initial state for a new
set of allowed tunneling transitions $|s_{i'} \rangle \to \{ |s_{f'} \rangle
\}$ (satisfying $|s_{f'} - s_{i'}| =  1/2$). Since this new set in general
differs from the previous set of transitions $|s_i \rangle \to \{ |f\rangle
\}$ allowed before the GSC, at $h_{ s_i, s_{i'}}$ one set of lines in the
tunneling spectrum ends and another begins.  A line from the former connects
continuously to one from the latter only if its final state $|f \rangle$ can
be reached from both $|s_i \rangle $ and $|s_{i'} \rangle$ [i.e.\ if $s_f -
s_i = - (s_f - s_{i'})$]; in this case, the two lines $|s_i \rangle \to |f
\rangle$ and $|s_{i'} \rangle \to |f \rangle$ join at $h_{ s_i, s_i'}$ via a
kink, since $\Delta E_{if} (h)$ and $\Delta E_{if'} (h)$ have slopes of
opposite sign.  However, for most lines this is not the case (since usually
$|s_f - s_{i'}| \neq  1/2$), so that at $h_{ s_i, s_i'}$ the line
$|s_{i}\rangle \to |f \rangle$ simply {\em ends}\/ while new lines $|s_{i'}
\rangle \to |f' \rangle$ begin.  This results in discontinuities (or
``jumps'') in the spectrum at $h_{ s_i, s_i'}$ of size $(\Delta E_{i'f'} -
\Delta E_{if})(h_{ s_i, s_i'})$, unless by chance some other final state $|f
'\rangle $ happens to exist for which this difference equals zero.

Since the order in which the GSCs $(s_i, s_{i'})$ occur as functions of
increasing $h$ depend on $d$ and $p$, as indicated by the distinct regimes I,
II, III, \ldots in Fig.~\ref{fig:h-crit}, one finds a distinct kind of
tunneling spectrum for each regime, differing from the others in the positions
of its jumps and kinks.  In regime~I, where the order of occurrence of GSCs
with increasing $h$ is $(0,1); (\frac12,\frac32); (1,2);
(\frac32,\frac52);\ldots$, there are no discontinuities in the evolution of
the lowest line [see Fig.~\ref{fig:spectra}(a)].  For example, for the $e\to
o$ spectrum, the lowest $|0 \rangle \to |1/2\rangle$ line changes {\em
  continuously\/} to $|1\rangle \to |1/2\rangle$ at $h_{0,1}$, since $|s_f -
s'_i | = 1/2$.  However, in all other regimes the first change in ground state
spin (at $h_{0, s_1}$ from 0 to $s_1$) is $ > 1$, implying a {\em jump}\/
(though possibly small) in all $e\to o$ lines, as illustrated by
Fig.~\ref{fig:spectra}(b).
 
The jump's magnitude for the tunneling thresholds, i.e.\ the lowest $e\to o$
and $o\to e$ lines, is shown as function of $d$ in Fig.~\ref{fig:jumps}.  It
starts at $d=0$ from the CC value $\tilde \Delta (1 - 1/\sqrt2)$ measured for
thin Al films,\cite{Meservey-94} and with increasing $d$ decreases to 0
(non-monotonically, due to the discrete spectrum).  This {\em decrease of the
  size of the jump in the tunneling threshold}\/ reflects the fact, discussed
in section~\ref{magfield}, that the change in spin at the first ground state
change $(s_0,s_1)$ decreases with increasing $d$ (as $s_1 \!-\! s_0 \sim
h_{CC}/d$), and signals the softening of the first-order
superconducting-to-paramagnetic transition.
 
The fact that the measured tunneling thresholds in
Fig.~\ref{fig:experimental-spectra} show no jumps at all (which might at first
seem surprising when contrasted to the threshold jumps seen at $h_{CC}$ in
thin films in a parallel field \cite{Meservey-70}), can therefore naturally be
explained \cite{Braun-97} by assuming the grain to lie in the ``minimal
superconductivity'' regime I of Fig.~\ref{fig:h-crit} (where the jump size
predicted in Fig.~\ref{fig:jumps} is zero).  Indeed, the overall evolution
(i.e.\ order and position of kinks, etc.)  of the lowest lines of
Fig.~\ref{fig:experimental-spectra} qualitatively agrees with those of a
regime I tunneling spectrum, Fig.~\ref{fig:spectra}(a).  This allows us to
deduce the following values for the level-crossing fields $H_{s_i,s'_i}$
(indicated by vertical dashed lines in Figs.~\ref{fig:experimental-spectra}
and~\ref{fig:spectra}): $H_{0,1} = 4$T, $H_{1/2,3/2} = 4.25$T, $H_{1,2} =
5.25$T and $H_{3/2,5/2} = 6.5$T. \label{HSS} As corresponding uncertainties we
take $\Delta H_{s_i,s'_i} = 0.13$T, which is half the $H$ resolution of 0.25T
used in experiment.

By combining the above $H_{s_i,s'_i}$ values with Fig.~\ref{fig:h-crit}, some
of the grain's less-well-known parameters can be determined somewhat more
precisely: Firstly, the grain's ``bulk $H_{CC}$'' field can be estimated by
noting from Fig.~\ref{fig:h-crit} that $h_{0,1} / h_{CC} \simeq 0.95$, so that
$H_{CC} = H_{0,1}/ 0.95 \simeq 4.2 T$.  This is in rough agreement with the
value $H_{CC} \simeq 4.7 T$ found experimentally\cite{Meservey-94} in thin
films in a parallel field, confirming our expectation that these correspond to
the ``bulk limit'' of ultrasmall grains as far as paramagnetism is concerned.
(Recall that our numerical choice of $\lambda = 0.224$ in
Section~\ref{generalnumerics} was based on this correspondence.)  Secondly,
the grain's corresponding bulk gap is $\tilde \Delta = \sqrt 2 \mu_B H_{CC}
\simeq 0.34$meV. Thirdly, to estimate the level spacing $d$, note that since
$H_{1/2,3/2} / H_{0,1} \simeq 1.06$, this grain lies just to the right of the
boundary between regions II and I in Fig.~\ref{fig:h-crit} where $d/ \tilde
\Delta \simeq 0.63$, i.e.\ $d \simeq 0.21$meV. (The crude volume-based value
$d \simeq 0.45$meV of Section~\ref{expdetails} thus seems to have been an
overestimate.)  It would be useful if the above determination of $d$ could be
checked via an independent accurate experimental determination of $d$ directly
from the spacing of lines in the tunnel spectrum; unfortunately, this is not
possible: the measured levels are shifted together by interactions, implying
that their spacing does not reflect the mean {\em independent}-electron level
spacing $d$.

The higher lines plotted in Fig.~\ref{fig:spectra} correspond to states where
the electron tunnels into an excited spin-$s_f$ state. For simplicity we
considered only excited states $|s_f, j \rangle$ involving a single
electron-hole excitation relative to $|s_f \rangle$, such as the example
discussed in section~\ref{excited} or as sketched in
Fig.~\ref{fig:alpha-states}(b), though in general others are expected to occur
too. The jumps in these lines (e.g. in Fig.~\ref{fig:spectra}(a) at $h_{1,2}$)
occur whenever the two final excited states $|s_f, j_f \rangle$ and $|s_{f'},
j_{f'} \rangle$ before and after the GSC at $h_{s_i, s'_i}$ have different
correlation energies.  (Recall that the correlation energy of an excited state
$|s_f, \bbalpha_f \rangle$ can be non-zero even if that of the corresponding
ground state $|s_f \rangle$ is zero, since the former's unpaired electrons are
further away from $\varepsilon_F$, so that
$\Delta_{s_f,\bbalpha_f}>\Delta_{s_f}$, see Section~\ref{excited}.)
Experimentally, these jumps have not been observed.  This may be because
up-moving resonances lose amplitude and are difficult to follow\cite{Ralph-97}
with increasing $h$, or because the widths of the excited resonances ($\simeq
0.13\tilde\Delta$) limit energy resolution.\cite{Agam-97}

For somewhat larger grains, the present theory predicts jumps even in the
lowest line.  It remains to be investigated, though, whether orbital effects,
which rapidly increase with the grain size, would not smooth out such jumps.

Finally, note that more than qualitative agreement between theory and
experiment can not be expected: in addition to the caveats mentioned in the
second paragraph of Section~\ref{generalnumerics}, we furthermore neglected
non-equilibrium effects in the tunneling process and assumed equal tunneling
matrix elements for all processes.  In reality, though, random variations of
tunneling matrix elements could suppress some tunneling processes which would
otherwise be expected theoretically.

\subsection{Parity Effects}

As mentioned in the introduction, several authors
\cite{vonDelft-96,Smith-96,Matveev-97,Braun-97} have discussed the occurrence
of a parity effect in ultrasmall grains: ``superconductivity'' (more
precisely, ground state pairing correlations) disappears sooner with
decreasing grain size in an odd than an even grain ($\Delta_{1/2} <
\Delta_{0}$, and $d_{c,1/2} < d_{c,0}$). This is a consequence of the blocking
effect, which is always stronger in the presence of an odd, unpaired electron
than without it.  This section is devoted to discussing to what extent this
and related parity effects are measurable.

Since pairing parameters such as $\Delta_{1/2}, \Delta_{0}$ are not observable
quantities, measurable consequences of parity effects must be sought in
differences between eigenenergies, which are measurable.

\subsubsection{$E_{1/2} - E_{0}$ is not measurable at present}
\label{Eoe}

One might expect that the odd-even ground state energy difference 
$E^{o/e}_G \equiv (E_{1/2} - E_0)$ should reveal traces of the parity
effect. Unfortunately, at present this quantity is not directly measurable,
for the following reasons: 

If the transport voltage $V$ is varied at fixed gate-voltage $V_g$, the energy
cost of changing the grain's electron number by one (the $h=0$ threshold
tunneling energy) depends [see Eq.~(\ref{energydiff})] not only on $E^{o/e}_G
$ but also on the change $\delta E_C$ in the grain's charging energy due to
tunneling.  However, as explained in Section~\ref{expdetails}, $\delta E_C$
depends (in an imprecisely-known way) on the actual value of $V_g$. Therefore
only the grain's {\em fixed}-$N$ excitation spectrum (distance between lines
of tunneling spectrum) can be measured accurately in this way, but not $
E^{o/e}_G$.

If the gate voltage $V_g$ is varied at a fixed transport voltage in the linear
response regime $V \simeq 0$, i.e.\ Coulomb oscillations are studied, one
expects to find a $2e$-periodicity in the so-called gate charge $Q_g = C_g V_g
+ const$, with $ E^{o/e}_G$ determining the amount of deviation from the
$e$-periodicity (see Appendix~\ref{sec:tichy} for details).  This has been
demonstrated convincingly in $\mu$m-scale devices \cite{Harvard,Saclay}.
However, for nm-scale devices it is at present not possible to study (as
suggested in Ref.~\onlinecite{Matveev-97}) $e-$ or $2e$-dependent features
with sufficient accuracy to carry through this procedure: the charging energy
is so large that the pairing-induced deviation from $e$-periodicity is a very
small effect (a fractional change of order $E_G^{o/e}/E_C < 0.01$), which can
easily be obscured by $V_g$-dependent shifts in background charge near the
transistor, as explained in Section~\ref{expdetails}.

\subsubsection{Parity Effect in Pair-Breaking Energies}
\label{sec:spectralgap}

Since the quantities that RBT can measure accurately are fixed-$N$ {\em
  excitation}\/ spectra, let us investigate what parity-effects can be
extracted from these.  Since any parity effect is a consequence
of the blocking effect, we begin by discussing the latter's most obvious
manifestation: it is simply the fact that breaking a pair costs correlation
energy, since the resulting two unpaired electrons disrupt pairing
correlations.  This, of course, is already incorporated in mean-field BCS
theory via the excitation energy of at least $2 \tilde \Delta$ involved in
creating two quasi-particles. It directly manifests itself in the qualitative
difference between RBT's even and an odd excitation spectra (explained in
section~\ref{expdetails}), namely that the former shows a large spectral gap
$2 \Omega_e >d$ between its lowest two lines that is absent for the latter
(Fig.~\ref{fig:experimental-spectra}).

The parity effect discussed by von Delft {\em et al.}\cite{vonDelft-96} and
Smith and Ambegaokar\cite{Smith-96} referred to a more subtle consequence of
the blocking effect that goes beyond conventional BCS theory, namely that the
pairing parameters $\Delta_s$ have a significant $s$-dependence once $d /
\tilde \Delta$ becomes sufficiently large.  Although these authors only
considered the ground state parity effect $\Delta_{1/2} < \Delta_{0}$, the
same blocking physics will of course also be manifest in generalizations to $s
> \frac12$.  In fact, the problems with measuring the odd-even ground state
energy difference $E^{o/e}_G$ discussed above leave us no choice but to turn
to $s > \frac12$ cases when looking for a measurable parity effect.
Specifically, we shall now show that a parity effect resulting from
$\Delta_{3/2} < \Delta_1$ should in principle be observable in present
experiments.

To this end, let us compare the $h=0$ {\em pair-breaking energies}\/
in an even and an odd grain, defined as  the energy per electron
 needed to break a
  single pair at $h=0$ by flipping a single spin: for an even grain, it
is $\Omega_e \equiv \frac12 (E_1 - E_0)_{h=0}$, i.e.\ simply half the spectral
gap discussed above; for an odd grain, it is $\Omega_o \equiv \frac12 (E_{3/2}
- E_{1/2})_{h=0}$.  

Within mean-field BCS theory, one would evaluate these using the same pairing
parameter $\tilde \Delta$ for all states and $[(\varepsilon_j - \mu_p)^2 +
\tilde \Delta^2]^{1/2}$ for the quasi-particle excitation energies associated
with having the single-particle state $|j, \pm \rangle$ definitely occupied or
empty, with parity-dependent chemical potential\cite{vonDelft-96} $\mu_p =
\varepsilon_0 - p d/2$. This would give $\Omega^{BCS}_e = [(d/2)^2 + \tilde
\Delta^2]^{1/2}$ and $\Omega^{BCS}_o = [d^2 + \tilde \Delta^2]^{1/2}$,
implying that the difference $\Omega^{BCS}_o - \Omega^{BCS}_e $ is strictly
$>0$ (Fig.~\ref{fig:spectral-gap}, dotted lines).  For $d/ \tilde \Delta \to
\infty$ this difference reduces to $d$, which is simply the difference in the
kinetic energy cost required to flip a single spin when turning
$|\frac12\rangle_0$ into $|\frac32\rangle_0$ (namely $2d$), relative to that
when turning $|0\rangle_0$ into $| 1\rangle_0$ (namely $d$).

In contrast, using the present theory to go beyond mean-field BCS theory, one
finds numerically that $\Omega_o > \Omega_e$ only for sufficiently large level
spacings ($d/ \tilde \Delta > 0.6$, see Fig.~\ref{fig:spectral-gap}, dotted
lines); for smaller $d$ one has $\Omega_o < \Omega_e$, implying that it costs
less energy to break a pair in an odd grain than an even grain, even though
the kinetic-energy cost is larger ($2 d$ vs.\ $d$). This happens since
$\Delta_{3/2} < \Delta_{1}$, which reflects a parity effect caused by
pair-blocking by the extra unpaired electron in $|3/2\rangle$ relative to
$|1\rangle$.  {\em The theoretical result that $\Omega_o / \Omega_e < 1$ for
  sufficiently small $d/ \tilde \Delta$ can be viewed as a ``pair-breaking
  energy parity effect''}\/ which is analogous to the ``ground state parity
effect'' $\Delta_{1/2} < \Delta_0$, but which, in contrast to the latter,
should be observable in the experimentally available {\em fixed}\/-$N$
eigenspectra.

What are $\Omega_e$ and $\Omega_o$ in RBT's experiments?
Unfortunately, the present data do not give an unambiguous answer: on
the one hand, the $h=0$ data allow the determination of
$\Omega_e=0.25$meV [half the $h=0$ energy difference between the two
lowest lines of Fig.~\ref{fig:experimental-spectra}(a)], but not of
$\Omega_o$, since breaking a pair is not the lowest-lying excitation
of an odd system at $h=0$ (which is why
Fig.~\ref{fig:experimental-spectra}(b) has no spectral gap).  On the
other hand, both $\Omega_e$ and $\Omega_o$ can be found from $h \neq
0$ data, since by Eq.~(\ref{eq:hcrit}) they are equal to the
level-crossing fields $h_{0,1}= \Omega_e$ and $h_{1/2,3/2}= \Omega_o$,
whose values were deduced from the experimental tunneling spectra in
Section~\ref{HSS}. This yields $ \Omega_e = 0.23\pm 0.01$meV and
$\Omega_o = 0.24 \pm 0.01$meV, i.e. a $\Omega_e$-value somewhat {\em
  smaller}\/ than the above-mentioned 0.25meV determined at $h=0$.
The reasons for this difference are presumably (i) that the actual
$g$-factors are not precisely 2 (as assumed), and (ii) that the
experimental spectral lines are not perfectly linear in $h$ (having a
small $h^2$-contribution due to orbital diamagnetism, neglected in our
model).

Nevertheless, if we assume that these two complication will not significantly
affect the ratio $ h_{1/2,3/2} / h_{0,1}$ (since $h_{1/2,3/2}$ and $h_{0,1}$
presumably are influenced by similar amounts), we may use it to estimate the
ratio $\Omega_o/\Omega_e = 4.25/4 = 1.06 \pm 0.1$.  This ratio is slightly
smaller than that expected from the mean-field BCS ratio $\Omega^{BCS}_o /
\Omega^{BCS}_e \simeq 1.1$ at $d /\tilde \Delta\simeq 0.63$, i.e.\ consistent
with the pair-breaking energy parity effect.  However, the difference between
1.06 and 1.1 is probably too small to regard this effect as having been
conclusively observed.

We suggest that it should be possible to conclusively observe the
pair-breaking energy parity effect in a somewhat larger grain with
$h_{1/2,3/2} < h_{0,1}$ (implying $\Omega_o/\Omega_e < 1$), i.e.\ in Regime~II
of Fig.~\ref{fig:h-crit}.  (This suggestion assumes that in regime II the
complicating effect of orbital diamagnetism is still non-dominant, despite its
increase with grain size.) To look for this effect experimentally would thus
require good control of the ratio $d / \tilde \Delta$, i.e.\ grain size.  We
suggest that this might be achievable if a recently-reported new fabrication
method, which allows systematic control of grain sizes by using colloidal
chemistry techniques\cite{Klein-97}, could be applied to Al grains.

\subsubsection{Parity Effect in the Limit $d / \tilde \Delta \gg 1$}

Since the parity effects discussed above are based on the observation that the
amount of pairing correlations, as measured by $\Delta_s$, have a significant
$s$-dependence, they by definition vanish for $d > d_{c,0}$, because then
$\Delta_s = 0$ for all $s$.  Matveev and Larkin (ML)\cite{Matveev-97} have
pointed out, however, that there is a kind of parity effect that persists even
in the limit $d/\tilde \Delta \gg 1$, which in the present theory we would
call the ``uncorrelated regime''
(since there the $\bar \Delta'$ defined in Eq.~(\ref{eq:v2u2})
would be $ \ll \tilde \Delta$):
 when one extra electron is added to an even
grain, it does not participate at all in the pairing interaction, simply
because this acts {\em only}\/ between pairs; but when another electron is
added so that now an extra pair is present relative to the initial even state,
it does feel the pairing interaction and makes a self-interaction contribution
$-\lambda d$ to the ground state energy.  To characterize this effect, they
introduced the pairing parameter
\begin{eqnarray}
  \Delta_P^{\rm ML} = E^{N+1}_{1/2}-
  {\textstyle \frac12}\left({E^{N}_0+E^{N+2}_0}\right),
\end{eqnarray}
with $N$=even. In first order perturbation theory in $\lambda$, i.e.\ using
$E^{N+p}_{p/2} \equiv {}_0\langle p| H | p \rangle_0$ (where $|p \rangle_0$ is
the uncorrelated Fermi ground state with $N+p$ electrons), one obtains
$\Delta_P^{\rm ML,pert} = \frac12 \lambda d$. This illustrates that this
parity effect exists even in the complete absence of correlations, and
increases with $d$.

Since our variational ground states $|p \rangle$ reduce to the uncorrelated
Fermi states $|p\rangle_0$ when $\Delta_P = 0$, the above perturbative result
for $d / \tilde \Delta$ can of course also be retrieved from our variational
approach: we approximate $E^{N}_0$ and $E^{N+1}_{1/2}$ by $\E_0(d)$ and
$\E_{1/2}(d)$, respectively, both of which were calculated above, and
$E^{N+2}_0$ by $\E_0(d)-\lambda d$, since it differs from $\E_0(d)$ only by an
extra electron pair at the band's bottom, whose interaction contribution in
Eq.~(\ref{eq:Esalpha}) is $-\lambda d (v_j^{(s)})^4$.  Thus the variational
result for  ML's parity parameter is
\begin{eqnarray}
  \label{MLvar}
  \Delta_P^{\rm ML,var} = \E_{1/2}(d)-\E_0(d)+ \lambda d / 2 \; ,
\end{eqnarray}
(see Fig.~\ref{fig:Matveev-parity}), which reduces to the 
perturbative result $ \Delta_P^{\rm ML,pert}$ for $d > d_{c,0}$. 
The reason why this parity effect did not
surface in the discussions of previous sections in spite of its linear
increase with $d$ is simply that there we were interested in {\em
  correlation}\/ energies of the form ${\cal E} - {\cal E}^0$ in which effects
associated with ``uncorrelated'' states were subtracted out [see e.g.\ 
Figs.~\ref{fig:pairing-parameter}(b) and \ref{fig:excited-states}(b)].

The perturbative result $\Delta_P^{\rm ML,pert} = \frac12 \lambda d$ is in a
sense trivial. However, ML showed that a more careful calculation in the
regime $d/ \tilde \Delta \gg 1$ leads to a non-trivial upward renormalization
$\tilde\lambda$ of the bare interaction constant $\lambda$,
\begin{equation}
  \label{eq:ren}
  \tilde\lambda = \frac{\lambda}{1-\lambda\log(\omega_C/d)} .
\end{equation}
To obtain $\Delta_P^{\rm ML}$  with logarithmic
accuracy,   $\lambda$ in  $ \Delta_P^{\rm ML,pert}$ is replaced 
by this renormalized $\tilde \lambda$, with the result
\begin{equation}
  \label{MLren}
  \Delta_P^{\rm ML} = d /(2 \log d/ \tilde \Delta).
\end{equation}
(The range of validity of this result lies beyond that shown in
Fig.~\ref{fig:Matveev-parity}, which therefore does not also show
$\Delta_P^{\rm ML}$.)  This logarithmic renormalization, which is beyond the
reach of our variational method (but was confirmed using exact diagonalization
in Ref.~\onlinecite{Fazio-97}), can be regarded as the ``first signs of
pairing correlations'' in what we in this paper have called the ``uncorrelated
regime'' [in particular since $|\Delta_P^{\rm ML}|$ {\em increases}\/ upon
renormalization only if the interaction is attractive, whereas it decreases
for a repulsive interaction, see Eq.~(\ref{eq:ren})].

Unfortunately, $\Delta_P^{\rm ML}$ is at present not measurable, for the same
experimental reasons as apply to $E^{e/o}_G$, see Section~\ref{Eoe}.

\subsection{Time Reversal Symmetry} 
\label{sec:time-reversed}

When defining our model in Eq.~(\ref{eq:hamilton}), we adopted a {\em
  reduced\/} BCS Hamiltonian, in analogy to that conventionally used for
macroscopic systems. In doing so, we neglected interaction terms of the form
\begin{equation}
\label{nontimereversed}
  -d \sum_{iji'j'} \lambda (i,j,i',j') c^\dagger_{i+}c^\dagger_{j-} 
  c_{i'-}c_{j'+} 
\end{equation}
between non-time-reversed pairs $c^\dagger_{i+}c^\dagger_{j-}$, following
Anderson's argument \cite{Anderson-59} that for a short-ranged interaction,
the matrix elements involving time-reversed states $c^\dagger_{j+}
c^\dagger_{j-}$ are much larger than all others, since their orbital
wave-functions interfere constructively\cite{Alsthuler-priv}.  Interestingly,
the experimental results provide strikingly direct support for the correctness
of neglecting interactions between non-time-reversed pairs of the from
(\ref{nontimereversed}) at $h=0$: Suppose the opposite, namely that the matrix
elements $\lambda (j+k,j, j'+ k, j')$ were all roughly equal to $\lambda$ for
a finite range of $k$-values (instead of being negligible for $k \neq 0$, as
assumed in $H_{\rm red}$). Then for $2s < k$, one could construct a spin-$s$
state $|s \rangle'$ with manifestly lower energy (${\cal E}'$) than that
(${\cal E}$) of the state $|s\rangle$ of Eq.~(\ref{eq:ansatzs}):
\begin{eqnarray}
\label{sprime}
    |s\rangle' = \!\!\!\! \prod_{j=-m}^{-m+2s-1} \!\!\!\! c^\dagger_{j+} 
              \!\!  \prod_{i= -m}^{\infty} (u^{(s)}_i \!+\! v^{(s)}_i
       c^\dagger_{(i+2s)+}c^\dagger_{i-})\,|\mbox{Vac}\rangle \; .
\end{eqnarray}
Whereas in $|s\rangle$ pair-mixing occurs only between time-reversed partners,
in $|s\rangle'$ we have allowed pair-mixing between {\em non\/}-time-reversed
partners, while choosing the $2s$ unpaired spin-up electrons that occupy their
levels with unit amplitude to sit at the band's {\em bottom} (see
Fig.~\ref{fig:repaired-state}).  To see that $|s\rangle'$ has lower energy
than $|s\rangle$,
\begin{equation}
\label{compareEss}
\label{eq:Esalphaprimed}
      {\cal E'}_s \; = \;{\cal E'}^{corr}_s \! + {\cal E'}^0_s 
      \; < \; {\cal E}^{corr}_s \! + {\cal E}^0_s \;= \; {\cal E}_s \; ,
\end{equation}
we argue as follows: Firstly, ${\cal E'}^0_s = {\cal E}^0_s $, since the
corresponding uncorrelated states $|s\rangle'_0$ and $|s \rangle_0$ are
identical [and given by Eq.~(\ref{eq:ansatzs00})].  Secondly, $\Delta'_s =
\Delta_0 (> \Delta_s)$, and hence ${\cal E'}^{corr}_s ={\cal E}^{corr}_0 (<
{\cal E}^{corr}_s \le 0)$, because the $2s$ unpaired electrons in $|s\rangle'$
sit at the band's bottom, i.e.\ so far away from $\varepsilon_F$ that their
blocking effect is negligible (whereas the $2s$ unpaired electrons in
$|s\rangle$ sit around $\varepsilon_F $ and cause significant blocking). Thus
Eq.~(\ref{compareEss}) holds, implying that $|s' \rangle $ would be a better
variational ground state for the interaction~(\ref{nontimereversed}) than $|s
\rangle$.

Now, the fact that ${\cal E'}^{corr}_s ={\cal E}^{corr}_0$ is {\em
  independent}\/ of $s$ means that flipping spins in $|s \rangle'$ does not
cost correlation energy. Thus, the energy cost for turning $|0 \rangle'$ into
$|1 \rangle'$ by flipping one spin is simply the kinetic energy cost $d$,
implying a threshold field $h'_{0,1} = d/2$ [see Eq.~(\ref{eq:hcrit})]; in
contrast, the cost for turning $|0 \rangle$ into $|1 \rangle$, namely $2
\Omega_e$, implies a threshold field $h_{0,1} = \Omega_e$, which (in the
regime $d \,{\stackrel{\scriptscriptstyle <}{\scriptscriptstyle \sim}} \,
\tilde \Delta$) is rather larger than $d/2$.  The fact that RBT's experiments
[Fig.~\ref{fig:spectra}(b)] clearly show a threshold field $h_{0,1}$
significantly larger than $d/2$ shows that the actual spin-$1$ ground state
chosen by nature is better approximated by $|1 \rangle$ than by $|1 \rangle'$,
in spite of the fact that ${\cal E'}_1 < {\cal E}_1 $.  Thus the premise of
the argument was wrong, and we can conclude that those terms in
Eq.~(\ref{nontimereversed}) not contained in $H_{red}$ can indeed be
neglected, as done in the bulk of this paper.

\section{Conclusions}

Citing the extensive literature in nuclear physics on fixed-$N$ projections of
BCS theory, we argued that a reasonable description of ultrasmall grains is
possible using grand-canonical BCS-theory, despite the fact that such grains
would strictly speaking require a canonical description.  Using a generalized
variational approach to calculate various eigenenergies of the grain, we
demonstrated the importance of the blocking effect (the reduction of
pair-mixing correlations by unpaired electrons) and showed that it becomes
stronger with decreasing grain size. The blocking effect is revealed in the
magnetic-field dependence of the tunneling spectra of ultrasmall grains, in
which pairing correlations can be sufficiently weak that they are destroyed by
flipping a single spin (implying ``minimal superconductivity'').  Our theory
qualitatively reproduces the behavior of the tunneling thresholds of the
spectra measured by Ralph, Black and Tinkham as a function of magnetic field.
In particular, it explains why the first order transition from a
superconducting to a paramagnetic ground state seen in thin films in a
parallel field is softened by decreasing grain size.  Finally, we argued that
a pair-breaking energy parity effect (that is analogous to the presently
unobservable ground state energy parity effect discussed previously) should be
observable in experiments of the present kind, provided the grain size can be
better controlled than in RBT's experiments.

{\em Acknowledgments:}\/ We would like to thank I.~Aleiner, B.~Altshuler,
V.~Ambegaokar, S.~Bahcall, C.~Bruder, D.~Golubev, B.~Janko, K.~Matveev,
A.~Rosch, A.~Ruckenstein, G.~Sch\"on, R.~Smith and A.~Zaikin for enlightening
discussions. Special thanks go to D.C.~Ralph and M.~Tinkham for a fruitful
collaboration in which they not only made their data available to us but also
significantly contributed to the development of the theory. This research was
supported by the German National Scholarship Foundation and the ``SFB~195'' of
the Deutsche For\-schungs\-ge\-mein\-schaft.

%
%
%

\def\f#1#2{\mbox{$\frac{#1}{#2}$}}

\begin{appendix}

\section{Analytical Limits}
\label{analytics}

\subsection{$d \to 0$  and Euler-MacLaurin Expansion}
\label{dto0}

When the level spacing $d$ tends to zero the theory reduces to the
conventional BCS variational and mean field approach. We can calculate the
properties of a superconducting system to first order in $d$ by expanding the
BCS solution around $d=0$. In doing so, we focus on the ground states
$|s\rangle$ of each spin-$s$ sector of Hilbert space. 

While in the bulk limit ($d=0$) the shift $-\lambda d(v_j^{(s)})^2$ in the
single-electron energies $\xi_j$ just after Eq.~(\ref{eq:gap}) is unimportant,
it influences the behaviour of an ultrasmall grain by effectively increasing
the level-spacing near the Fermi surface. Its effect is largest for $s=0$,
since for $s \neq 0$ the states at the Fermi surface, where the deviation of
$v_j^{(s, \alpha)}$ from 0 or 1 is largest, are blocked. For simplicity
we neglect the $v_j^{(s,\alpha)}$-dependence in $\xi_j$ in the following
calculation, using $\xi_j = \varepsilon_j-\mu - \lambda d \,
\theta(-(\varepsilon_j-\mu))$, and therefore good agreement with numerics can
only be expected for $d\ll\tilde\Delta$ and $s \neq 0$. Within this
approximation for $\xi_j$, $\mu$ lies halfway between the topmost double
occupied and lowest completely empty level in $|s\rangle_0$: $\mu =
\varepsilon_0 - d(\delta_{p,0}+\lambda)/2$. Note that $\mu$ does {\em not} lie
exactly on one of the levels in the odd case ($p=1$) as one might have
expected at first sight, but 
halfway between the topmost doubly-occupied and lowest completely
empty level.

\label{sec:smallDelta}
We shall calculate the pairing parameter $\Delta_s(d)$ in the small-$d$ limit
by calculating the first terms of its  Taylor series:
\begin{eqnarray}
\label{expandD}
  \Delta_s(d) & \simeq & (1+d \partial_d +\frac{d^2}{2} \partial^2_d)
  \Delta_s(0)   \; .
\end{eqnarray}
To this end, it suffices to solve the gap equation (\ref{eq:gap}), as well its
first and second derivatives with respect to $d$, for $d=0$. This can be done
by rewriting Eq.~(\ref{eq:gap}) using the Euler-MacLaurin summation formula,
\begin{eqnarray}
  \label{eq:Euler-MacLaurin}
\lefteqn{1/ \lambda= d \sum_{j=j_0}^{j_1}  
f(j d) \simeq 
     \int_{ j_0 d}^{j_1 d} d\xi\,f(\xi)} \\ &&
+\frac{d}2[f(j_0 d )+f( j_1 d )]+  
 \frac{d^2}{12}[f'(j_0 d)+f'(j_1 d)],\nonumber
\end{eqnarray}
with $f(j d)=[(jd )^2+\Delta_s^2]^{-1/2}$, $j_0 = s + (1+ \lambda)/2$ and $j_1
= \omega_c/d$. 
The $s$-dependence has now been absorbed in the lower bound $j_0$ of the
sum. The negative branch of the sum is identical to the positive since $\mu$
lies halfway between the topmost doubly-occupied and lowest completely empty
level. 
It therefore suffices to calculate the positive branch times two.
Setting $d=0$ in Eq.~(\ref{eq:Euler-MacLaurin}) yields the well-known BCS bulk
gap equation, whose solution is, by definition, $\Delta_s(0) = \tilde\Delta$.
The first and second total $d$-derivatives of Eq.~(\ref{eq:Euler-MacLaurin})
yield
\begin{math}
  \partial_d \Delta_s (d=0) = -s
\end{math}
and $\partial_d^2 \Delta_s (d=0)=-s^2/\tilde\Delta$, so that
the desired result from Eq.~(\ref{expandD}) is
\begin{eqnarray}
  \label{eq:gap-s}
  \Delta_s(d) & \simeq & \tilde\Delta-{(s+\lambda/2)d} - \frac{(s+\lambda/2)^2
  d^2}{2\tilde\Delta}.
\end{eqnarray}

We next calculate the eigenenergies ${\cal E}_s$ by evaluating
Eq.~(\ref{eq:Esalpha}) up to first order in $d$, where the sums again are
evaluated with the help the the Euler-MacLaurin formula.  Since we are
interested in the effects of pairing correlations we subtract the energy
$\E^0_p$ of the uncorrelated Fermi sea $|p\rangle_0$:
\begin{eqnarray}
  \label{eq:Es-small-d}
  \begin{array}{lrcl}
  \mbox{(even)} & \E_s-\E^0_0 & \simeq &        
     -\frac{\tilde\Delta^2}{2d}+(1+\frac{\pi}4)\lambda
     \tilde\Delta+2s\tilde\Delta-\\
& & & {}-(s^2-\frac1{12}+\frac{\pi+6}4\lambda s d;\\
  \mbox{(odd)}  & \E_s-\E^0_{\frac12} & \simeq & 
   -\frac{\tilde\Delta^2}{2d}+\frac{\pi}4\lambda\tilde\Delta+2s\tilde\Delta-{}\\
& & & {}-(s^2+\frac16+\frac{\pi+6}4\lambda s+\frac\lambda2) d.
  \end{array}
\end{eqnarray}
The $d^{-1}$ term is the bulk correlation energy, which is slightly
renormalised by the intensive $(1+\frac\pi4)\lambda\tilde\Delta$-term, which
in turn stems from the $v^4$-terms of Eq.~(\ref{eq:Esalpha}). $2s\tilde\Delta$
is the bulk excitation energy for $2s$ quasi-particles. The $d^1$-term is the
first-order correction for discrete level-spacing.

\subsection{$d$ near $d_c$ and the Small Delta Expansion}
\label{sec:neardc}

The other analytically tractable limit is $d\gg\Delta_s$, which holds for $d$
near the critical spacing $d_{c,s}$ where $\Delta_s$ vanishes.

First, we derive an expression for the critical $d_{c,s}$ by solving the gap
equation with vanishing pairing parameter $\Delta_s$ for $d$:
\begin{eqnarray}
  \frac1{\lambda} = \sum_{j=j_0}^{\omega_c/d_{c,s}} \frac1{j} =
  \Psi(\omega_c/d_{c,s}+1) - \Psi(j_0).
\end{eqnarray}
$\Psi(x)$ denotes the Digamma function and $j_0$ equals $s+\f{1+\lambda}2$
again.  Remembering that $\lambda = 1/\log(\frac{2\omega_c}{\tilde\Delta})$
and $\exp(\Psi(x)) \sim x-\f12$ for large $x$ this equation reduces to
\begin{eqnarray}
  \log\left(\frac{2d_{c,s}}{\tilde\Delta}\right) & = & -\Psi(s+\f{1+\lambda}2) \\
  d_{c,s} & = & \frac{\tilde\Delta}2 \exp(-\Psi(s+\f{1+\lambda}2)).
\end{eqnarray}
For $s\ge1$ this can be simplified to
\begin{eqnarray}
  d_{c,s} \simeq \frac{\tilde\Delta}{2s+\lambda}.
\end{eqnarray}
Numerical values of $d_{c,s}/\tilde\Delta$ ($\lambda=0.224$) are $2.36, 0.77,
0.44, 0.31,\ldots$ for $s=0, \frac12, 1, \frac32, \ldots$ respectively. Near
$d_{c,s}$ the pairing parameter vanishes like
\begin{eqnarray} \label{eq:square-root}
  \Delta_s \simeq \tilde\Delta\sqrt{1-\frac{d}{d_{c,s}}}\quad\mbox{for
    $d\gg\Delta_s$ and $s>0$},
\end{eqnarray}
which we shall now show.

Since for the spin-$s$ ground states with vanishing pairing parameter electron
and hole pairs are symmetrically distributed around the Fermi surface,
Eq.~(\ref{eq:mu}) again yields $\mu=\varepsilon_0 - d
({\delta_{p,0}+\lambda}/2)$.  We turn to the gap equation (\ref{eq:gap}). The
spin dependence has been absorbed in $j_0$. The positive and negative branches
of the restricted sum are identical (because of the special symmetric value of
$\mu$), with $|\xi|$ ranging from $d(s+\frac{1+\lambda}2) = dj_0$ to
$\omega_c$. It therefore suffices to calculate the positive branch times two:
\begin{eqnarray}
  \frac{1}{\lambda} & = & \sum_{j=j_0}^{\omega_c/d}
     \left(j^2+\Delta_s^2/d^2\right)^{-1/2}  \nonumber\\
\label{gap22}
 & \simeq & \sum_{j=s+(1+\lambda)/2}^{\omega_c/d}
     \left(\frac1{j}-\frac{\Delta_s^2}{2d^2j^3}\right) \\
 \sum_{j=s+(1+\lambda)/2}^{\omega_c/d_{c,s}} \frac1{j} & \simeq & 
 \sum_{j=s+(1+\lambda)/2}^{\omega_c/d}
     \left(\frac1{j}-\frac{\Delta_s^2}{2d^2j^3}\right) \nonumber
\end{eqnarray}
To obtain Eq.~(\ref{gap22}),
the square root was
expanded using $\Delta_s\ll d$.
The remaining sums can be expressed by the polygamma
functions $\Psi^{(n)}$ using the identity
\begin{eqnarray}
  \sum_{k=1}^n \frac1{k^m} = \zeta(m) - (-1)^m \frac{\Psi^{(m-1)}(n+1)}{(m-1)!}.
\end{eqnarray}
Replacing the sums by the Polygamma functions and collecting terms leads to
\begin{eqnarray}
  \Psi\left(\frac{\omega_c}{d_{c,s}}+1\right) -
  \Psi\left(\frac{\omega_c}{d}+1\right)  = \nonumber\\ 
  \hfill {} -\frac{\Delta_s^2}{4d^2}
  \left[\Psi''\left(\frac{\omega_c}{d}+1\right) -  
    \Psi''\left(s+\frac{1+\lambda}2\right)\right] .
\end{eqnarray}
Now assume that $d$ is close to $d_{c,s}$: $d=d_{c,s}-\delta d$ and $\delta d
\ll d_{c,s}$. Expand the left hand side in $\delta d$ and use the asymptotics
for $\Psi'$ (on the left hand side) and $\Psi''$ (on the right hand side) for
the large $\omega_c/d$ argument. Also the $\Psi''(s+\frac12)$-term is
approximated by its asymptotic form $-s^{-2}$:
\begin{eqnarray}
  \label{eq:noasympt}
  \frac{\delta d}{d_{c,s}} & = & -\frac{\Delta_s^2}{4d^2}\Psi''(s+\f12) \\
  \Delta_s^2           & = & 4d^2s^2 \frac{d_{c,s}-d}{d_{c,s}} \\
  \Delta_s             & = & \tilde\Delta\sqrt{1-\frac{d}{d_{c,s}}}.
\end{eqnarray}
The last step was performed by remembering that $4d^2s^2 = 4d_{c,s}^2s^2 \simeq
\tilde\Delta^2$ for $s\neq0$. 

Although Eq.~(\ref{eq:square-root}) was derived for $d$ near $d_{c,s}$, it
turns out to have a surprisingly large range of validity: its small-$d$
expansion in powers of $d/\tilde \Delta$ agrees (at least) up to second order
with Eq.~(\ref{eq:gap-s}), and for $s\ge1$ it in fact excellently reproduces
the numerical results for $\Delta_s(d)$ for {\em all}\/ $d$.

For $s=0$ the asymptotic expansion of $\Psi''$ breaks down. Therefore directly
from (\ref{eq:noasympt}) we deduce
\begin{eqnarray}
  \Delta_0 \simeq \sqrt{\frac{4d^2}{12.1}\frac{d_{c,s}-d}{d_{c,s}}},
\end{eqnarray}
where we used  $\Psi''(\frac{1+\lambda}2)\simeq-12.1$. This result gives
good agreement with numerics
near $d_{c,s=0}$, but obviously  has the wrong $d\to0$ limit.

\section{$I$-$V$ characteristics of an ultrasmall NSN SET}
\label{sec:tichy}

The $I$-$V$ characteristics of a SET with an ultrasmall superconducting grain
as island, i.e.\ an ultrasmall NSN SET, were examined by Tichy and von
Delft\cite{Tichy-96}.  They described the discrete pair-correlated eigenstates
of the grain using the parity-projected mean-field BCS theory of
Ref.~\onlinecite{vonDelft-96}.  Although this approach is too crude to
correctly treat pairing correlations of excited states (since for all even (or
odd) ones the {\em same}\/ $\Delta_0$ (or $\Delta_{1/2}$) is used), it does
treat the even and odd ground states correctly.  It therefore enables one to
understand how the odd-even ground state energy difference $E^{o/e}_G \equiv
(E_{1/2} - E_0)$ should influence the SET's $I$-$V$ characteristics.

\begin{figure}
  \begin{center}
    \leavevmode \epsfxsize 7cm
    \epsffile{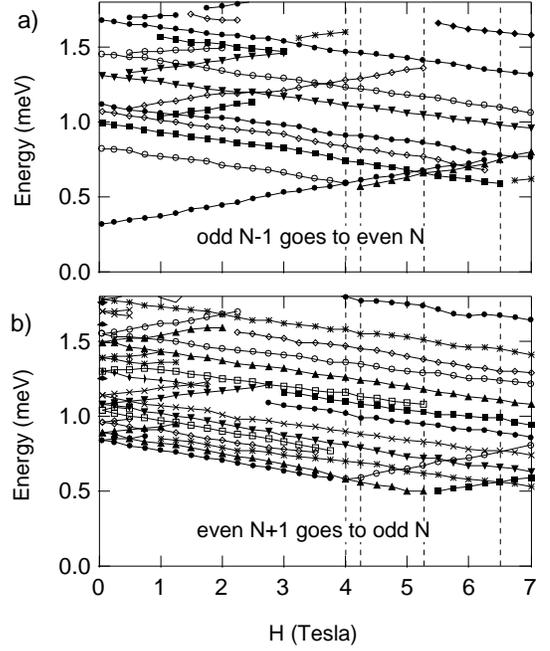}
    \vspace*{5mm} \caption{Experimental tunneling spectra measured by
      RBT (Fig.~3 of Ref.~\protect\onlinecite{Ralph-97}).  The
      distances between lines give the fixed-$N$ excitation spectra of
      (a) an even and (b) an odd grain, as explained in
      Section~\ref{expdetails}.  The vertical dashed lines indicate
      the first four level-crossing fields $H_{s,s'}$ (assigned by
      comparison with Fig.~\ref{fig:spectra}, see
      Section~\ref{sec:tunneling-spectra}), namely $H_{0,1} \!=\!4$T,
      $H_{1/2,3/2}\!=\!4.25$T, $H_{1,2}\!=\!5.25$T and
      $H_{3/2,5/2}\!=\!6.5$T with uncertainty $\pm 0.13$T (half the
      $H$-resolution of 0.25T).}
    \label{fig:experimental-spectra} \vspace*{5mm}
  \end{center}
\end{figure}

\begin{figure}
  \begin{center}
    \leavevmode \epsfxsize 8cm
    \epsffile{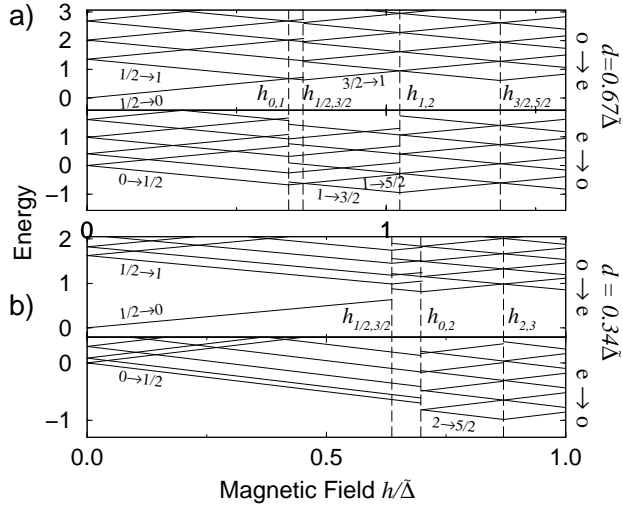}
  \end{center}
  \vspace*{5mm} \caption{The theoretical odd-to-even and even-to-odd
    tunneling spectra $(\Delta E_{if} - \Delta E_{min} (0))/ \tilde
    \Delta $ predicted for an ultrasmall superconducting grain as a
    function of magnetic field $h$, for two different level spacings:
    (a) $d=0.67\tilde\Delta$ and (b) $d=0.34\tilde\Delta$
    (corresponding to regimes I and III of
    Fig.~\protect\ref{fig:h-crit}, respectively).  Some lines are
    labeled by the corresponding $s_i\to s'_i$ tunneling transition.
    Not all possible higher lines (corresponding to excited final
    states $|s,j\rangle$) are shown.  Vertical dashed lines indicate
    those level-crossing fields $h_{s,s'}$ [see Eq.~(\ref{eq:hcrit}]
    at which kinks or jumps occur, with $h_{0,1}<h_{1/2,3/2}< h_{1,2}
    < h_{3/2,5/2}$ in (a) and $h_{1/2,3/2}<h_{0,2} < h_{2,3}$ in (b).}
  \label{fig:spectra} \vspace*{5mm}
\end{figure}

\begin{figure}
  \begin{center}
    \leavevmode \epsfxsize 7cm
    \epsffile{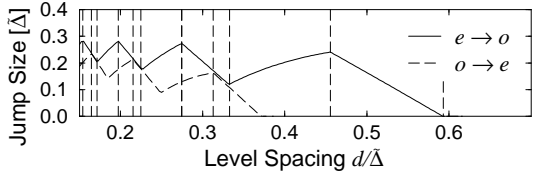}
    \vspace*{5mm} \caption{The first jump in the lowest line of a tunneling
      spectrum occurs at the level-crossing field $h_{CC}(p,d)= h_{s_0,s_1}$
      that induces the first ground state change $(s_0,s_1)$. The size of this
      jump in units of $\tilde \Delta$, namely $|\Delta E_{s_1,f'} - \Delta
      E_{s_0,f}|/ \tilde \Delta$, is shown here as function of $d/
    \tilde \Delta $ for $e\to
      o$ (solid line) and $o \to e$ (dashed line) tunneling spectra.  Both
      lines approach the CC value $1 - 1/\sqrt2 = 0.29$ as $d \to 0$.  Their
      non-monotonic behavior is due to the discreteness of the level spacing;
      the kinks occur at the regime boundaries of Fig.~\ref{fig:h-crit},
      indicated here by vertical dashed lines.}
    \label{fig:jumps} \vspace*{5mm}
  \end{center}
\end{figure}

\begin{figure}
  \begin{center}
    \leavevmode \epsfxsize 8cm
    \epsffile{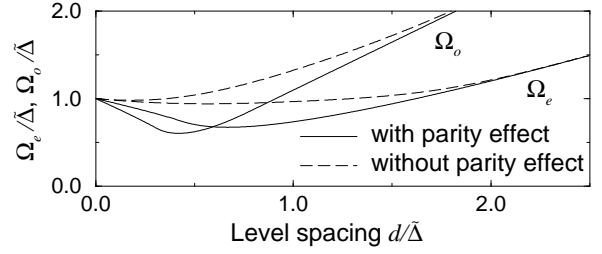}
  \end{center}
  \vspace*{5mm} \caption{Parity effect for the pair-breaking energies
    $\Omega_e \equiv \frac12 (E_1 - E_0)_{h=0}$ and $\Omega_o \equiv
    \frac12 (E_{3/2} - E_{1/2})_{h=0}$ (see
    Section~\ref{sec:spectralgap}): when calculated naively using
    conventional mean-field theory (dashed lines), the pair-breaking
    energies obey $\Omega_o > \Omega_e$ for all $d / \tilde \Delta$;
    in contrast, when calculated within generalized variational BCS
    theory (solid lines), $\Omega_o < \Omega_e$ for $d / \tilde \Delta
    < 0.6$; this reflects a parity effect, namely that $\Delta_{3/2} <
    \Delta_{1}$, which is caused by the extra unpaired electron in
    $|3/2\rangle$ relative to $|1\rangle$.}
  \label{fig:spectral-gap} \vspace*{5mm}
\end{figure}

\begin{figure}
  \begin{center}
    \leavevmode \epsfxsize 8cm
    \epsffile{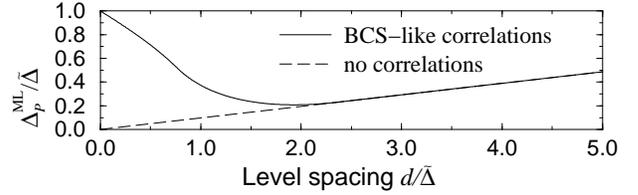}
    \vspace*{5mm} \caption{The parity parameter $\Delta_P^{\rm ML}$ discussed
      by Matveev and Larkin \protect\cite{Matveev-97}, calculated perturbatively
      for the uncorrelated Fermi sea
      ($\Delta_P^{\rm ML,pert}= \frac12 \lambda d$, dashed line), and using our
      generalized variational BCS-approach ($\Delta_P^{\rm ML,var}$ of
      Eq.~(\protect\ref{MLvar}), solid line). The renormalized result
      $\Delta_P^{\rm ML}$ of Eq.~(\protect\ref{MLren}) given by ML is not
      shown since its range of validity is $d / \tilde \Delta \gg 1$.}
    \label{fig:Matveev-parity} \vspace*{5mm}
  \end{center}
\end{figure}

\begin{figure}
  \begin{center}
    \leavevmode \epsfxsize 8cm
    \epsffile{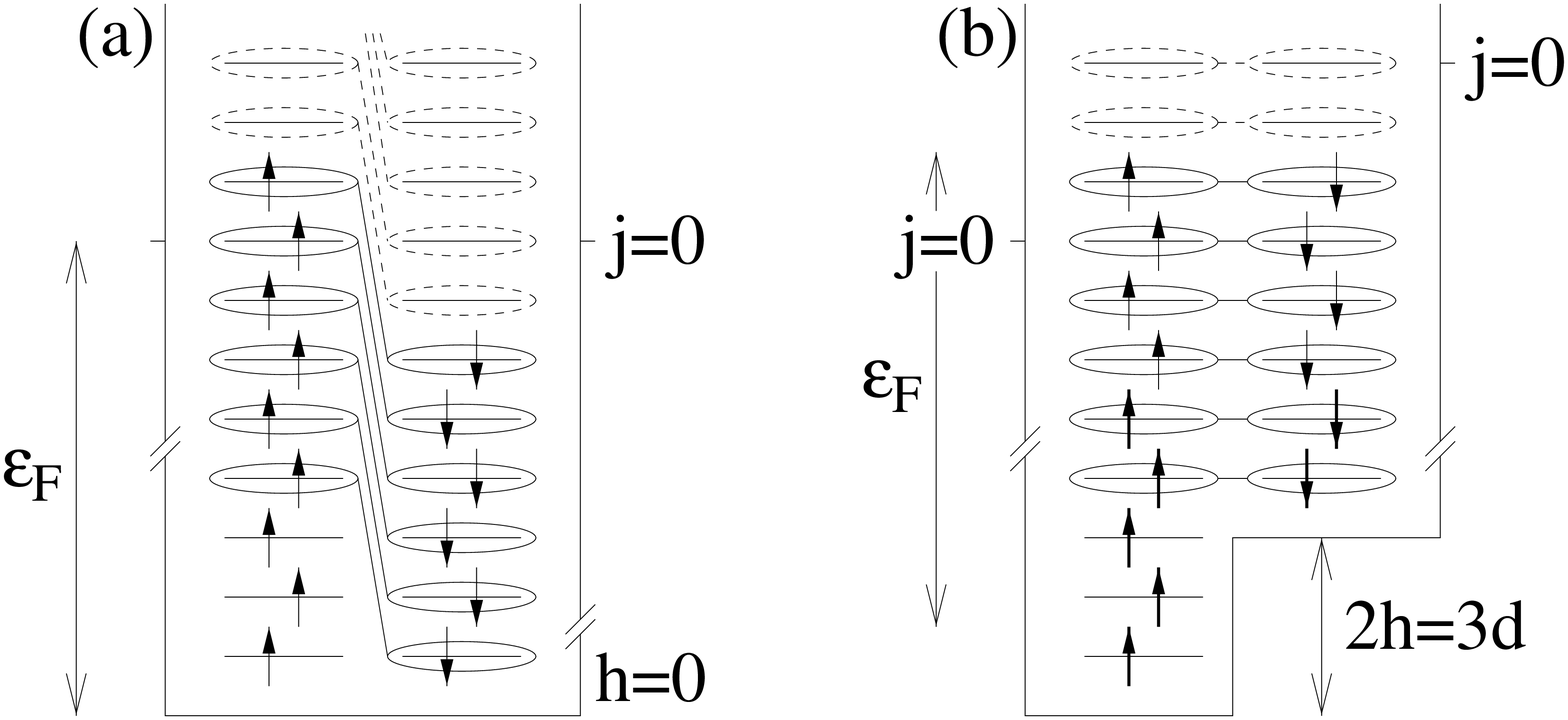}
    \vspace*{5mm} \caption{Schematic representations of the
      non-time-reversed-pairing state $|3/2 \rangle'$ defined in
      Eq.~(\ref{sprime}).  The energies $\varepsilon_{j} \mp h$ of the
      single-particle states $|j, \pm \rangle$ are drawn (a) for $h=0$ and (b)
      for $2h = 3d$. We indicated schematically how non-time-reversed
      states are paired according to $(u_i + v_i c^\dagger_{(i+3)+}
      c^\dagger_{i-})$ in the BCS-like Ansatz~(\ref{sprime}),
      with solid or dashed ellipses encircling states that would be
      completely filled or empty in the absence of pairing correlations.}
   \label{fig:repaired-state} \vspace*{5mm}
  \end{center}
\end{figure}

\begin{figure}[htbp]
  \begin{center}
    \leavevmode \epsfxsize 8cm
    \epsffile{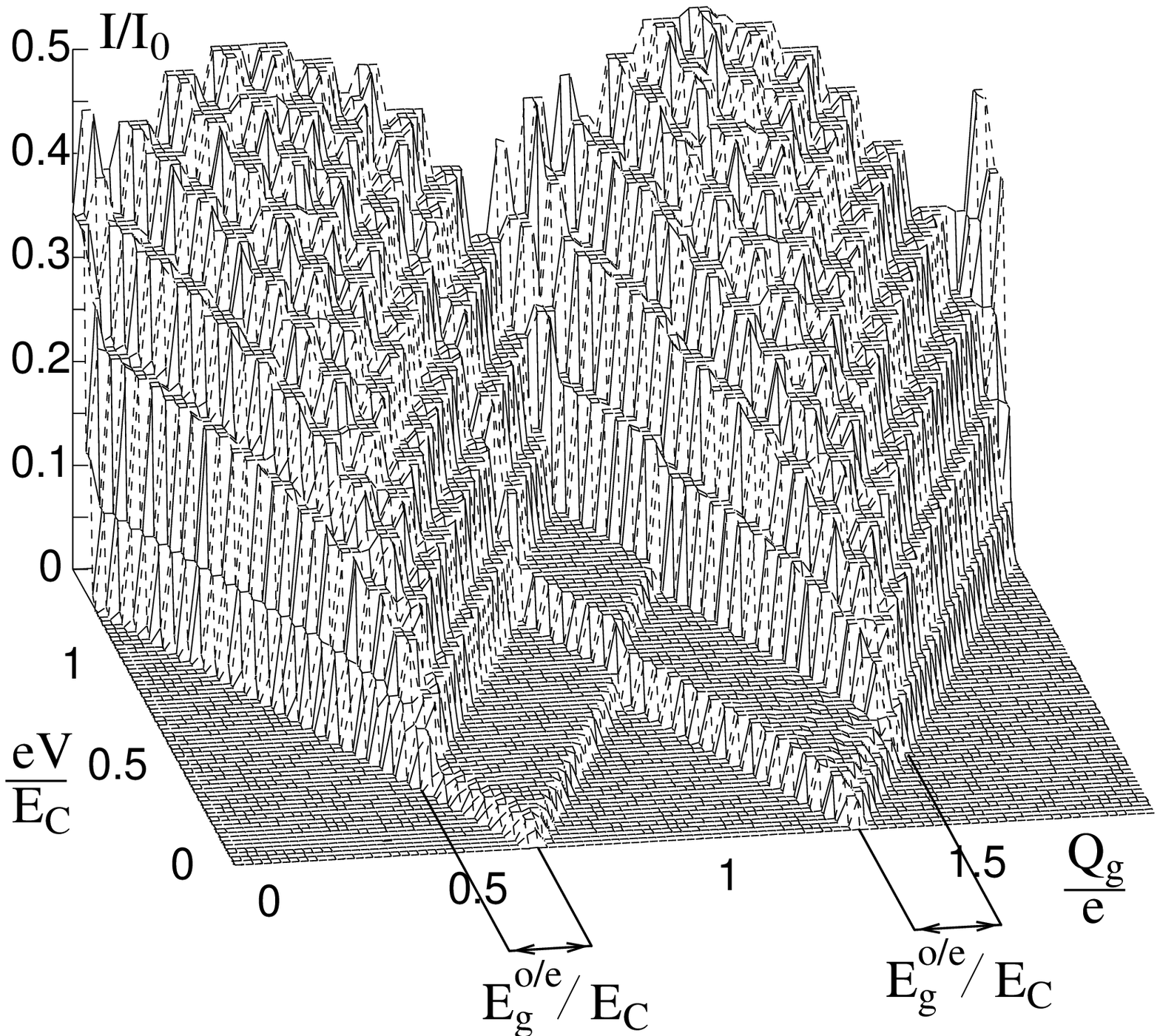}
    \vspace*{5mm} \caption{$I$-$V$ characteristics for a SET with an
      ultrasmall superconducting grain as island (from
      Ref.~\protect\onlinecite{Tichy-96}).  The current is plotted as a
      function of gate charge ($Q_g/e$) and transport voltage ($eV/E_C$).
      Pairing correlations shift the degeneracy-point values of $Q_g/e$ away
      from their $e$-periodic values of $2m \pm \frac12$ by $\pm
      E^{o/e}_G/E_C$ (see text). To better reveal the figure's characteristic
      features, it was plotted using a ratio $E^{o/e}_G/E_C \simeq 0.1$, very
      much larger than the typical values of $< 0.01$.}
    \label{fig:current} \vspace*{5mm}
  \end{center}
\end{figure}

Using tunneling rates given by Fermi's golden rule and solving an appropriate
master equation, Tichy calculated the tunnel current through the SET as a
function of transport voltage $V$ and gate voltage $V_g$ at zero magnetic
field. In an ideal sample, the $I$-$V$ characteristics are $2e$-periodic in
the gate charge $Q_g = V_g C_g + const$; one such period is shown in
Fig.~\ref{fig:current}.  The usual Coulomb-blockade ``humps'' centered roughly
around the degeneracy points $Q_g/e = 2m \pm \frac12$ are decorated by
discrete steps, due to the grain's discrete eigenspectrum.  In RBT's
experiments $V_g$ was fixed near a degeneracy point and the current measured
as function of $V$ (for a set of different $H$-values).  When following a line
parallel to the $V$-axis in Fig.~\ref{fig:current}, the positions of the steps
in the current thus correspond to the $H=0$ eigenenergies of RBT's tunneling
spectra in Fig.~\ref{fig:experimental-spectra}.

The reason for $2e$- instead of $e$-periodicity are pairing correlations:
Firstly, the grain's odd-even ground state energy difference $E^{o/e}_G$
causes a shift in the degeneracy-point values for $Q_g/e$ from $2m \pm
\frac12$ to $2m \pm (\frac12+ E^{o/e}_G/E_C)$.  Secondly, tunneling spectra
measured in the $V$-direction in Fig.~\ref{fig:current} show a plateau after
the first step if the final state after tunneling is even (i.e.\ for $\frac12
+ E^{o/e}_G/E_C \le Q_g / e \le \frac32 -E^{o/e}_G/E_C $), but not if it is
odd, corresponding to the presence or absence of a large spectral gap in the
tunneling spectra of Fig.~\ref{fig:experimental-spectra}(a) or (b); this is
due to the energy cost to break a pair, and the plateau's width is simply
twice the even pair-breaking energy $2\Omega_e$ (see
Section~\ref{sec:spectralgap}).

\end{appendix}

\newpage

\newpage

\newpage

\newpage

\begin{thebibliography}{10}
  
\bibitem{Anderson-59} A.W. Anderson, J.\ Phys.\ Chem.\ Solids {\bf 11}, 28
  (1959).

\bibitem{GiaverZeller-68} I. Giaver and H. R. Zeller, Phys.\ Rev.\ Lett. {\bf
    20}, 1504 (1968).

\bibitem{ZellerGiaver-69} H. R. Zeller and I. Giaver, Phys.\ Rev. {\bf 181},
  789 (1969).

\bibitem{Strongin-70} M. Strongin, R. S. Thompson, O. F. Kammerer, and J. E.
  Crow, Phys.\ Rev.\ B {\bf 1}, 1078 (1970).
  
\bibitem{Muehlschlegel-72} B. M{\"u}hlschlegel, D.J. Scalapino, and R. Denton,
  Phys.\ Rev.\ B {\bf 6}, 1767 (1972).

\bibitem{Ralph-95} D. C. Ralph, C. T. Black, and M. Tinkham, Phys.\ Rev.\ 
  Lett. {\bf 74}, 3241 (1995).

\bibitem{Black-96} C. T. Black, D. C. Ralph, and M. Tinkham, Phys.\ Rev.\ 
  Lett. {\bf 76}, 688 (1996).
  
\bibitem{Ralph-96B} D. C. Ralph, C. T. Black, and M. Tinkham, Physica
  (Amsterdam), {\bf 218B}, 258 (1996).

\bibitem{Ralph-97} D. C. Ralph, C. T. Black, and M. Tinkham, Phys.\ Rev.\ 
  Lett. {\bf 78}, 4087 (1997).
  
\bibitem{Agam-97} O. Agam, N. S. Wingreen, B. L. Altshuler, D. C. Ralph, and
  M. Tinkham, Phys.\ Rev.\ Lett. {\bf 78}, 1956 (1997).
  
\bibitem{AgamAleiner-97} O. Agam and I. L. Aleiner, Phys.\ Rev.\ B {\bf 56},
  5759 (1997).

\bibitem{vonDelft-96} J. von Delft, A. D. Zaikin, D. S. Golubev, and W. Tichy,
  Phys.\ Rev.\ Lett. {\bf 77}, 3189 (1996).
  
\bibitem{Smith-96} R. A. Smith and V. Ambegaokar, Phys.\ Rev.\ Lett. {\bf 77},
  4962 (1996).
  
\bibitem{Matveev-97} K. A. Matveev and A. I. Larkin, Phys.\ Rev.\ Lett. {\bf
    78}, 3749 (1997).
  
\bibitem{Braun-97} F. Braun, J. von Delft, D. C. Ralph, and M. Tinkham,
  Phys.\ Rev.\ Lett. {\bf 79}, 921 (1997).
  
\bibitem{Bahcall-priv} S. Bahcall, unpublished.

\bibitem{Janko-94}
B. Jank\'o, A.\ Smith, and V.\ Ambegaokar,  
Phys.\ Rev.\ B {\bf 50}, 1152 (1994).


\bibitem{Golubev-94}
D. S. Golubev and A. D.Zaikin, 
Phys.\ Lett.\ A {\bf 195}, 380   (1994); {\em Qantum Dynamics of
Submicron Structures}\/, H. A. Cerdeira {\em et al.}\/ (Eds.)
Kluwer Acad. Publ., 473 (1995). 

\bibitem{Soloviev-61} V. G. Soloviev, 
  Mat.\ Fys.\ Skrif.\ Kong.\ Dan.\ Vid.\ Selsk. {\bf 1}, 1 (1961).
  
\bibitem{Blanter-96} Ya. M.  Blanter, private communication (1996).

\bibitem{Meservey-94}
R. Meservey and P. M. Tedrow,  Phys.\ Rep. {\bf 238}, 173 (1994).

\bibitem{RingSchuck-80} P. Ring and P. Schuck, {\em The Nuclear
    Many-Body Problem}, Springer-Verlag, (1980).

\bibitem{BCS-57} L. N. Cooper, J. Bardeen, and J. R. Schrieffer, Phys.\ Rev.
  {\bf 108}, 1175 (1957).
  
\bibitem{Ralph-priv} D. C. Ralph. \newblock Private communication.
  
\bibitem{Braun-98} F. Braun, and J. von Delft, to be published.

\bibitem{Balain-97} R. Balain, H. Flocard, and M. Veneroni, nucl-th/9706041
  (1997).

\bibitem{Fazio-97} A. Mastellone, G. Falci, and R. Fazio, unpublished (1997). 

\bibitem{Garland-68} J. W. Garland, K. H. Bennemann, and F. M. Mueller, Phys.\
  Rev.\ Lett. {\bf 21}, 1315 (1968).

\bibitem{Meservey-70} R. Meservey, P. M. Tedrow, and Peter Fulde, Phys.\ Rev.\ 
  Lett. {\bf 25}, 1270 (1970).

\bibitem{Chandrasekhar-62} B. S. Chandrasekhar, Appl.\ Phys.\ Lett. {\bf 1}, 7
  (1962).

\bibitem{Clogston-62} A. M. Clogston, Phys.\ Rev.\ Lett. {\bf 9}, 266 (1962).

\bibitem{Harvard} M. T.  Tuominen, J. M.  Hergenrother, T. S. Tighe, and M.\ 
  Tink\-ham, Phys.\ Rev.\ Lett. {\bf 69}, 1997 (1992); Phys.\ Rev.\ B {\bf
    47}, 11599 (1993); M. Tinkham, J. M. Hergenrother, and J. G. Lu, Phys.\ 
  Rev.\ B {\bf 51}, 12649 (1996).
  
\bibitem{Saclay} P. Lafarge, P. Joyez, D. Esteve, C. Urbina, and M.H. Devoret,
  Phys.\ Rev.\ Lett.\ {\bf 70}, 994 (1993).

\bibitem{Klein-97} D. L. Klein, R. Roth, A. K. L. Lim, A. P. Alivisatos and
  Paul L.  McEuen, Nature {\bf 389}, 699 (1997).

\bibitem{Alsthuler-priv} Boris L. Altshuler, private communication (1997).
  
\bibitem{Tichy-96} Wolfgang Tichy,  Diploma thesis, University Karlsruhe (1996).
  
\end{thebibliography}
\end{document}